\pdfoutput=1

\documentclass[twocolumn,nofootinbib,showpacs,preprintnumbers,amsmath,amssymb]{revtex4-1}
\usepackage{graphicx}% Include figure files \usepackage{dcolumn}% Align table columns on decimal point \usepackage{bm}% bold math 
\usepackage{epsfig}
\newcommand{\be}{\begin{equation}} \newcommand{\ee}{\end{equation}} 
\newcommand{\ba}{\begin{eqnarray}} \newcommand{\ea}{\end{eqnarray}}

\newcommand{\bea}{\begin{eqnarray}}
\newcommand{\eea}{\end{eqnarray}}
\begin{document}
\preprint{
      OHSTPY-HEP-T-10-003\cr
      NSF-KITP-10-10-053\cr
     UCB-PTH-10/10\cr}
 \vspace{-1cm}
  \vspace*{\stretch1}
\newcommand{\we}{\wedge}
\title{An M Theory Solution to the Strong CP-problem, and Constraints on the Axiverse}
\author{Bobby Samir Acharya\footnote{bacharya@cern.ch}}
\affiliation{Abdus Salam International Centre for Theoretical
Physics, Strada Costiera 11, Trieste, Italy, and\\INFN, Sezione di Trieste and\\
MCTP, University of Michigan, Ann Arbor, MI 48109, USA}
\author{Konstantin Bobkov\footnote{bobkov@mps.ohio-state.edu}}
\affiliation{Department of Physics\\ The Ohio State University, Columbus, OH 43202, USA, and\\
Kavli Institute for Theoretical Physics, Kohn Hall, UCSB,  Santa Barbara, CA 93106, USA}
\author{Piyush Kumar\footnote{kpiyush@berkeley.edu}}
\affiliation{Berkeley Center for Theoretical Physics
University of California, Berkeley, CA 94720, and\\
Theoretical Physics Group
Lawrence Berkeley National Laboratory, Berkeley, CA 94720}
\vspace{0.5cm}

%\date{\today}

%\vspace{0.3cm}

\begin{abstract}
We give an explicit realization of the ``String Axiverse" discussed in Arvanitaki et. al \cite{Arvanitaki:2009fg} by extending our previous results on
moduli stabilization in $M$ theory to include axions. We extend the analysis of \cite{Arvanitaki:2009fg} to allow for high scale inflation that leads to a moduli dominated pre-BBN Universe. We demonstrate that an axion which solves the strong-CP problem naturally arises and that both the axion decay constants and GUT scale can consistently be around $2\times 10^{16}$ GeV with a much smaller fine tuning than is usually
expected. Constraints on the Axiverse from cosmological observations, namely isocurvature perturbations and tensor modes are described.
Extending work of Fox et. al \cite{Fox:2004kb}, we note that {\it the observation of tensor modes at Planck will falsify the Axiverse completely.}
Finally we note that Axiverse models whose lightest axion has mass of order $10^{-15}$ eV and with decay constants
of order $5\times 10^{14}$ GeV require no (anthropic) fine-tuning, though standard unification at $10^{16}$ GeV is difficult to accommodate.

\end{abstract}
\maketitle
%\newpage
%\vspace{-1.2cm} 

\section{Introduction} \label{intro}

The dimensionless QCD $\theta$-angle is a source of CP violation in the Standard
Model, highly constrained by measurements of the electric dipole moment (EDM) of the
neutron and $^{199}Hg$: $|\theta_{qcd}| \lesssim
10^{-10}$ \cite{Baker:2006ts}. This is the strong
CP problem, {\it why is $\theta_{qcd}$ so small?}.

An elegant solution to the problem might be provided by
the Peccei-Quinn (PQ) mechanism \cite{Peccei:1977hh}, in which
$\theta_{qcd}$  is promoted to a dynamical field known as the QCD axion ($a_{qcd}\equiv
f_a^{qcd}\,\theta_{qcd}$).
The axion is charged under an anomalous global $U(1)$ symmetry, broken by non-perturbative QCD
effects at the mass-scale $f_a^{qcd}$.
The potential generated by the QCD instantons is: \ba \label{Vqcd}V \sim
\Lambda_{qcd}^4\,\left[1-\cos\left(\frac{a_{qcd}}{f_a^{qcd}}\right)\right]\,,\ea which is minimized at
$\theta_{qcd} = 0$, solving the strong CP problem. Various properties of the QCD axion have been considered in \cite{oldaxion}, \cite{Raffelt}.

However, in order to really solve the problem, this PQ symmetry must be of incredibly high quality. In particular, all contributions to the axion potential from other PQ breaking sources must be at least ten orders of magnitude suppressed compared to that from QCD instantons.  Phenomenological approaches in field theory trying to realize such a PQ symmetry have been pursued
\cite{axionpheno}; see \cite{Cheung:2010hk} for a recent approach. However, since global symmetries are believed to be always broken by quantum gravitational effects, it is worthwhile exploring axions in string theory and under what circumstances they could solve the strong-CP problem. 

Pseudoscalar fields with axion-like properties generically arise in string/$M$ theory as the zero modes of antisymmetric tensor fields along the extra
dimensions \cite{Witten:1984dg}. 
The number of axions is determined essentially by the topology of the extra dimensions. This number, like the gauge group and number of fermion generations, can be viewed as
a {\it discrete},  UV boundary condition and
it is fairly common to have hundreds, if not thousands, of axions present. There is a separate
PQ symmetry for each axion field, inherited from the gauge symmetry associated with the higher dimensional tensor field.
Often these axions pair up with geometric moduli fields to form complex chiral mutliplets in ${\cal N}=1,D=4$ supergravity, the effective theory describing  many string compactifications.
Although a plethora of axions occur in string/$M$ theory, they do not automatically satisfy the criteria required by the PQ mechanism for the QCD axion. The absence of PQ breaking operators is usually guaranteed to all orders in perturbation theory because 
the PQ symmetries are remnants of (higher dimensional) gauge symmetry, surviving as global symmetries below the Kaluza-Klein scale. 
As in ordinary QCD, this global symmetry can be broken by non-perturbative effects, such as those arising from instantons of various sorts - worldsheet, brane, gauge and gravitational. One has to ensure that in addition to being valid to all orders in perturbation theory, the contributions to the QCD axion potential from these non-perturbative effects are negligible compared to that from QCD instantons. Axions in various string theoretic contexts have been studied in \cite{axionstring}.

A qualitative picture of the mass spectrum of axions in string theory can be given.
Since the potential of a given axion $a$ is generated by non-perturbative effects, its mass-squared will be of order $\frac{\Lambda^4}{f^2_a}$ -- where
$\Lambda^4 \sim M^4 e^{-bS_{inst}}$, where $S_{inst}$ is the action of the instanton,  which dominates the potential of $a$ and $b$ is a number,
which characterises the "charge of the instanton" that generated the potential \cite{Svrcek:2006yi}. $M$ is given by the geometric mean of the supersymmetry breaking scale ($\sqrt{F}$) and the `fundamental' scale, eg. the Planck scale. We will derive a precise formula later with the above general form. $S_{inst}$ is usually the volume of a submanifold of the extra dimensions in fundamental units. We will
later see that $f_a$ is typically of order the $GUT$ scale.
Therefore, the spectrum of axion masses will typically be exponentially hierarchical. Roughly speaking, with a large number of axions, one expects
a spectrum, which is uniformly distributed on a log scale, rather like the Yukawa couplings in $M$ theory \cite{Acharya:2003gb}. The axion masses are
thus expected to span many orders of magnitude.
These sorts of observations inspired the authors of \cite{Arvanitaki:2009fg} to propose a variety of astrophysical tests, which could probe the
large range of axion masses from
$10^{-33}$eV to $10^8$ eV. These tests could provide evidence for the existence of an Axiverse.
These include observations of the polarization of the CMB, suppression of the matter power spectrum at small scales
and the spectrum of gravitational waves from rapidly rotating astrophysical black holes. 
The purpose of this paper is not to discuss these phenomena and tests in detail,
but to address in more depth the nature of the mass spectrum of axions in string/$M$ theory and the solution to the strong CP problem. At the end, however, we will summarize some of the important astrophysical observables, which could either completely falsify the framework or provide strong evidence for it.

The requirement of stabilizing moduli with a sufficiently large mass in realistic string compactifications complicates the above picture for generating masses for the axions. For example, simple mechanisms of stabilizing moduli with fluxes in a supersymmetric minimum can completely break the PQ symmetry and give axions masses comparable to that of the moduli
\cite{Acharya:2002kv}. Stabilizing certain moduli by non-perturbative effects in a supersymmetric minimum, such as in the KKLT idea \cite{Kachru:2003aw}, also gives axion masses comparable to that of moduli; hence these axions \emph{cannot} solve the strong CP-problem. In fact, it can be shown that there does not exist any \emph{supersymmetric} minimum within ${\cal N}=1,D=4$ supergravity with phenomenologically allowed values of stabilized moduli but unfixed axions (down to the QCD scale), which could 
solve the strong CP problem \cite{Conlon:2006tq}. In particular, for supersymmetric vacua it was shown that either all moduli appear in the superpotential in which case the axions are very massive, or some moduli do not appear in the superpotential in
which case the potential is tachyonic \cite{Conlon:2006tq}. 

Therefore, in order to look for QCD axion candidates in string theory, one should consider
compactifications in which moduli are stabilized in a {\it
non-supersymmetric} minimum. Such vacua have been studied in detail in the context of
$M$ theory compactifications on $G_2$ manifolds 
without fluxes \cite{Acharya:2006ia, Acharya:2008hi}. In these compactifications \emph{all} the axions pair up with geometric moduli fields in such a way that all the moduli superfields enjoy a PQ symmetry, implying that the entire moduli superpotential can only arise from non-perturbative effects. In a series of papers \cite{Acharya:2006ia, Acharya:2008hi} it has been proven that strong dynamics in the hidden sector can a) generate a potential, which stabilizes {\it all} the moduli fields and b) generates a hierarchically small supersymmetry breaking scale.
Moreover, the potential stabilizes just one out of the many
axion fields. Happily,
the QCD axion belongs to the set of unfixed combinations. Higher order non-perturbative effects, which are generically
present, but not considered in \cite{Acharya:2006ia, Acharya:2008hi} because they are sub-dominant,  
{\it will} then stabilize the remaining axions (including the would-be QCD axion) at an exponentially smaller scale than the moduli mass scale (which
is set by the gravitino mass $m_{3/2}$). In other words, one considers the moduli to be stabilized at the  supersymmetry breaking scale $(m_{3/2})$ with sub-leading non-perturbative effects being responsible for stabilizing most of the axions.
Being non-perturbative, the resulting spectrum of axions will be distributed (roughly) uniformly on a logarithmic scale - as we indeed verify in a detailed model - thereby giving a detailed realization of the Axiverse. It can be shown that certain classes of Type IIB compactifications inspired by the moduli stabilization mechanism in $M$ theory \cite{Acharya:2006ia, Acharya:2008hi} also exhibit the above properties \cite{Bobkov:2010rf}. \emph{Therefore, the analysis and results of this paper can be directly applied to the those Type IIB compactifications as well}. For concreteness, however, we will study the effective theory arising from $M$ theory compactifications here.

The axion decay constants $f_a$ are constrained by various observations.
The decay constant for the QCD axion $f_a^{qcd}$ must be bounded from below by $\sim 10^9$ GeV because of processes like axion emission from stars and supernovae. 
An upper bound on {\it all} axions arises from the requirement of `not overclosing the Universe'. The thermal population of light axions ($\lesssim 1$ eV for $f_a \sim 10^9$ GeV) is typically quite negligible. However, being extremely weakly coupled to matter, axion fields begin coherent
oscillations, which begin when the Hubble scale $H \sim m_a$, and the constraint arises from not storing too much energy density in the axion fields. This energy density is quadratic in the initial vev (called mis-alignment angle) of the axion field. 
The value of the upper bound obtained clearly depends on the detailed cosmological evolution i.e. matter vs. radiation domination and entropy releases during the oscillations. 
Assuming a cosmology characterized by a radiation-dominated phase after inflation, one gets $f_a^{qcd} \lesssim 10^{12}$ GeV (the bound for general axions
is given in section IV.)
Thus, in order to satisfy the upper bound on $f_a^{qcd}$  \emph{without} fine-tuning the intial misalignement angle $\theta_i^{qcd}$, one has to either invoke a very large cycle on which the QCD gauge group is localized, and/or have a sufficiently low string/11D
scale, implying a large overall volume of the extra dimensions ${\cal V}$) \cite{Svrcek:2006yi}. 
Having a very large QCD cycle to make $f_a$ respect the
above upper bound makes the strong gauge coupling $\alpha_{QCD}$ far too weakly
coupled at the compactification scale and hence does not seem very natural.
So the only reasonable option without fine-tuning the misalignment angle, is to have a sufficiently small string/11D scale $M_s/M_{11} \ll M_{GUT}$ with a very large overall volume ${\cal V}$ but a small (still within the supergravity regime) QCD cycle such that $\alpha_{QCD}$ is reasonable \cite{Svrcek:2006yi}. However, this implies that the standard unification scale
$\sim 10^{16}$ GeV is above the string scale. Also, a low fundamental scale will likely lead to conflicts with current bounds on the proton lifetime.
Of course, it is possible that the initial misalignment angle 
is small ($\ll 1$) for some reason, which can make a large unification scale ($\sim 10^{16}$ GeV) compatible with 
the observational bound on $f_a$, assuming standard thermal cosmology. In fact, it has been argued that a small misalignment angle consistent with observations could be anthropically selected \cite{Linde:1991km}, \cite{Hertzberg:2008wr}. The $M$ theory models, as we will see, (and as is assumed in \cite{Arvanitaki:2009fg}) have $f_a \sim M_{GUT} \sim 10^{16}$ GeV.

One of the key points of \cite{Acharya:2008bk} is that, if the Hubble scale after inflation $H_I \geq m_{3/2} \sim 50$ TeV then the energy density of the Universe is dominated by moduli oscillations
until just before BBN. This is a different cosmological history than the one that yields the upper limit $f_a \leq 10^{12}$GeV. Moduli with
masses of order $50$ TeV will decay just before BBN and,
as pointed
out in \cite{Fox:2004kb}, the entropy released by the moduli decay can allow a larger $f_a \leq 5 \times 10^{14}$ GeV without fine tuning. 
Thus, the anthropic fine tuning of the initial
axion mis-alignment angle is much smaller since the upper bound is much closer to the GUT scale.  
\cite{Arvanitaki:2009fg} do not consider such moduli dominated cosmological histories (which are presumably generic in string theory)
and in fact take the Hubble scale to be
rather small after inflation $H_I \leq 0.1$GeV. It is somewhat satisfying that the fine-tuning of the misalignment angles is greatly reduced in the (presumably) more generic case of moduli dominated, non-thermal cosmology. 
We thus extend the results of \cite{Fox:2004kb} to the Axiverse and extend the Axiverse to the presumably more generic case of $H_I \geq m_{3/2}$. We find that the non-thermal moduli dominated pre-BBN Universe is much less fine tuned than a standard radiation dominated pre-BBN Universe arising in low scale inflation. Extending the results of \cite{Fox:2004kb}, which only considers the QCD axion, to the Axiverse, we emphasise that the observations of tensor modes by Planck, when combined with the existing bounds on axion induced isocurvature perturbations, would rule out the string Axiverse completely, thereby requiring axion masses to be `large'.

The plan of the paper is as follows. In section \ref{stable}, the mechanism to stabilize the axions is outlined in detail and the spectrum of eigenvalues and eigenstates is computed. A simple formula for the axion masses is obtained.
In section \ref{qcdaxion}, the effects of QCD instantons are taken into account and the precise method of identifying the QCD axion candidate among all axions is explained. It is shown that such a candidate typically occurs  within the $M$ theory framework considered. Section \ref{cosmo} is an analysis of the axion relic abundances within the `thermal' and `non-thermal' cosmological histories, followed by a discussion of the allowed parameter space after imposing current cosmological constraints in section \ref{Cons}. This is followed in  section \ref{observables} by a discussion of cosmological observables, which could either falsify or provide strong evidence for the framework, and also distinguish among the two cosmological histories. Technical details follow in Appendices \ref{masses}, \ref{toy} and \ref{stats}.

\section{Stabilization of Axions}\label{stable}

In this section we will describe the details of how, adding higher order non-perturbative corrections to the model considered in \cite{Acharya:2006ia,  Acharya:2008hi},
one can stabilize all the axion fields and calculate their masses. We work within the framework of supergravity in four dimensions with a superpotential $W$,
which is a holomorphic function of the moduli superfields $z_m \equiv t_m + is_m$ whose real parts are the axion fields $t_m$ and imaginary parts are the geometric
moduli $s_m$. There is a real Kahler potential $K$ depending on all of the $s_m$ and whose form is given below.

The superpotential we consider takes the form: 
\ba \label{W} W = A_1\phi_1^a\,e^{ib_1\,F_1}+A_2\,e^{ib_2\,F_1}+\sum_{K=3}^{\infty}\,A_K\,e^{ib_K\,F_K}\,,\ea

where all the coefficients $A_{K}$'s are order one {\it constants} (because of the PQ symmetries). 
The first two terms come from the model considered in \cite{Acharya:2006ia, Acharya:2008hi} and arise from strong gauge dynamics in the hidden sector. They depend on only one combination of axions. The remaining axions enter through the higher order corrections represented in the sum. These higher order corrections owe their origin to non-perturbative effects such as membrane instantons or gaugino condensates and are expected to be generically present in these compactifications. Note that in principle this is an infinite sum if one takes multiply wrapped instantons into account. In practice, though, one only requires considering as many independent terms as there are axions present. If the number of supersymmetric three-cycles in the compactification is sufficiently large, it is easy to have the required number of independent terms. Even if the number of supersymmetric three-cycles is not large enough one expects non-BPS instantons to contribute to the K\"{a}hler potential, which     
would give rise to the same results qualitatively; hence for simplicity and concreteness we assume henceforth that there are enough independent terms present in the superpotential above.

The $F$'s are integer linear combinations of the moduli superfields  ($F_K=\sum_{i=1}^N\,N_K^iz_i$) and
$\phi_1$ is a holomorphic composite field made of hidden sector matter fields. 
$a=-\frac{2}{P_1}$, $b_{1,2}=\frac{2\pi}{P_{1,2}}, P_{1,2}\in \bf{Z}^+$ are the dual Coxeter numbers of the hidden sector gauge theory, whereas
$b_3$, $b_4, ... = 2\pi\,I,\,I\in  \bf{Z}^+$ -- since the higher order terms are {\it assumed} to be generated by membrane instantons. This is consistent with these terms being higher order since
$V_K \equiv \rm{Im}(F_K) \geq 1$ is the volume of a three dimensional submanifold of the extra dimensions in 11d units. 
  
There are $N$ geometric axions $t_i\equiv {\rm Re}(z_i)$ and one matter axion $\theta$ (the phase of $\phi_1$ in (\ref{W})). It is also possible to consider a more general case with matter axions in the subdominant terms in (\ref{W}). We do not consider such a case below for simplicity. With the superpotential (\ref{W}) and a generic K\"{a}hler potential of the form: 
\ba 
\label{kahler} K=-3\log(\cal{V})+...,
\ea 
where $\cal{V}$ is the overall volume of the internal manifold in 11D units, the scalar potential contains $N$ approximately flat directions corresponding to $N$ PQ symmetries, which are preserved by the first two terms in (\ref{W}), while fixing the axion combination \cite{Acharya:2006ia}: 
\ba \label{co}
\cos(\chi_1 - \chi_2)\equiv \cos((b_1 - b_2)\vec{N_1}\cdot\vec{t}+a\theta)=-1\,,
\ea 
where $\chi_1 = b_1\vec{N}_1\cdot\vec{t}+a\theta,\,\chi_K = b_K \vec{N}_K\cdot\vec{t};\, K=2,3,..$.

With all moduli and one combination of axions fixed by the supergravity potential coming from the first two terms, the remaining $N$ would-be Goldstone bosons
of the PQ symmetries are fixed by the next $N$ largest terms in the potential, which contain $N$ linearly independent combinations of axions. This is because the terms in the potential proportional to $e^{-b_1 V_1 -b_MV_M}\cos(\chi_1-\chi_M)$ and $e^{-b_2 V_1-b_K V_K}\cos(\chi_2-\chi_K)$ where $M,\,K>2$ are much greater than in 
$e^{-b_M V_M -b_K V_K}\cos(\chi_M-\chi_K)$. Therefore, these terms fix not only the independent axion combinations
$\chi_1-\chi_M$ and $\chi_2-\chi_K$ but also effectively fix the combination $\chi_M-\chi_K$, as $\chi_M-\chi_K=(\chi_M-\chi_1)+(\chi_1-\chi_2)+(\chi_2-\chi_K)$ with $\chi_1-\chi_2$ fixed by (\ref{co}). 

\subsection{Axions - Mass Eigenvalues and Eigenstates}\label{spectra}

In order to compute the axion spectrum, it is best to study the effective potential for \emph{light} axions generated by integrating out the moduli and the heavy axion combination. Doing that gives rise to the following: 
\ba\label{veff}
V_{eff}&\approx& V_0-m_{3/2}m_{pl}^3\,e^{K/2}\sum_{K=3}^{N+2}D_K\,e^{-b_KV_K}\cos(\chi_1-\chi_K)
\nonumber\\&\forall k:&\,\,\,b_KV_K<b_{K+1}V_{K+1}\,,
\ea where $D_K$ are ${\cal O}(1)$ positive numbers not important for the order of magnitude estimates, $m_{3/2}\equiv e^{\langle K\rangle/2}\,\langle W\rangle\,m_{pl}$ and $V_K$ is the stabilized volume of the cycle that generated the corresponding term. 

After canonically normalizing the axion kinetic terms by taking the non-trivial K\"{a}hler metrics $\tilde{K}_{ab}\equiv\frac{\partial^2\,K}{\partial z_a\partial z_b};z_a\equiv\{s_i,\phi_1^0\}$ for moduli and matter fields into account, one finds the following estimate for the light axion masses (see Appendix \ref{masses}): \ba \label{mass-spec} 
\hat{m}_{a_i}^2\approx \frac {m_{3/2}m^3_{pl}}{\hat{f}_{a_{i+2}}^2}\,e^{K/2}c_{i+2}\,e^{-b_{i+2}V_{i+2}};\,i=\overline{1,N}\,,\ea where 
$c_{k}$ are ${\cal O}(1)$ model-dependent coefficients and the axion decay constants are given by $\frac{\hat{f}_{a_K}}{m_{pl}}=\sqrt{2\tilde{K}_{k}}$, with $\tilde{K}_{k}\delta_{km}\equiv U^{\dag}_{kn}\tilde{K}_{nl}U_{lm}$ the \emph{diagonal} K\"{a}hler metric obtained from the original K\"{a}hler metric by a unitary transformation $U$. In the notation above, the heavy axion combination stabilized with a mass comparable to the moduli is denoted by $\hat{m}_{a_0}$, hence the index $i$ labeling the light axions goes from 1 to $N$.

%A more schematic, but more widely applicable formula is
%\be
%m_{a_i} \sim {m_{pl} \over M_{GUT}} (m_{3/2}m_{pl})^{1/2}e^{-b_{i+2} V_{i+2}/2}
%\ee
%Thus, one only needs to know how the volumes of cycles are distributed in order to estimate the masses.

The results of \cite{Acharya:2006ia, Acharya:2008hi} show that all geometric moduli $s_j$ are stabilized in terms of only one parameter - $\langle V_1 \rangle$, 
which is the stabilized volume of the three-cycle supporting the hidden sector giving rise to the first two terms. This is the inverse of the coupling constant in the hidden sector $\langle V_1 \rangle = 1/\alpha_{hid}$.
$\langle V_1 \rangle$ is itself determined from $b_1$ and $b_2$ in (\ref{W}) \cite{Acharya:2006ia, Acharya:2008hi}. Since $V_K\equiv \sum_{j=1}^N\,N_K^j\,s_j$ and the overall volume ${\cal V}$ are determined in terms of $s_j$,  this implies that all three-cycle volumes $V_K$ and the overall volume ${\cal V}$ are stabilized as well. In particular, \ba\label{Vk} \langle V_K\rangle = \frac{3}{7}\,\langle V_1\rangle\,\sum_j\,\frac{N_K^j}{N_1^j}\tilde{a}_j\,,\ea where $N^j_K$, $N^j_1$ are integers and $\tilde{a}_j$ are real numbers \cite{Acharya:2008hi}.

It is convenient to express the masses of the light axions in terms of the gravitino mass. Using the result for $m_{3/2}$ in terms of $m_{pl}$ from \cite{Acharya:2006ia, Acharya:2008hi}
\ba \label{m32}
m_{3/2}\equiv e^{K/2}\,\langle W\rangle\,m_{pl}\approx e^{K/2}\Big |\frac{b_2-b_1}{b_1}\Big |\,A_2\,e^{-b_2\langle V_1\rangle}m_{pl}\,,
\ea 
one obtains the more widely applicable formula: 
\ba \label{massspec}
m_{a_i} = O(10^{-3})\,{m_{pl} \over M_{GUT}} (m_{3/2}m_{pl})^{1/2}e^{-b_{i+2} V_{i+2}/2}\,,
\ea 
where we have used $\hat{f}_{a_K}\approx M_{GUT}$ and ${\cal V}\sim 5000$ \cite{Acharya:2008hi}. 
We have also checked that the above result (\ref{massspec}) gives a very good estimate of the axion masses found exactly by numerical methods for a toy example with four axions (see Appendix \ref{toy}).

Since the integer coefficients in (\ref{Vk}) generically differ by ${\cal O}(1)$ for different three-cycles $V_K$, the masses will be distributed evenly on a logarithmic scale. Hence, \emph{this provides a concrete realization of the ``string Axiverse" considered in \cite{Arvanitaki:2009fg} with a multitude of extremely light axions ($\hat{m}_i \lll m_{3/2} \sim$ TeV) in a controlled and reliable manner}. The implications of this will be discussed in more detail later in the paper.  

What can be said about the range of the axion masses possible within this framework? 
This range is determined by the allowed range of $\{V_J\};\,J=\overline{3,N}$. For concreteness we assume that
all the $b_{I+2}=2\pi$.
To estimate the range of allowed values of $V_J$, it is useful to note the result of \cite{Friedmann:2002ty} 
where it was argued that the quantity $\eta_J \equiv \frac{{\cal V}}{4\pi^{1/3}V_J^{7/3}}$ is bounded from above by ${\cal O}(1)$ for realistic compactifications. In many cases, it can be much smaller than unity, i.e. $\eta_J \ll 1$. 

The upper bound $\eta_J={\cal O}(1)$ gives a lower bound on the three-cycle volume $V_J$, $\{V_{J}\}^{min}\approx (\frac{\cal V}{4\pi^{1/3}})^{3/7}$. From \cite{Acharya:2008hi}, the overall volume $\frac{{\cal V}}{4\pi^{1/3}}$ is generically stabilized between $\approx 550$ and $\approx 1150$ (in 11D units), leading to $15\lesssim \{V_{J}\}^{min} \lesssim 20$. Using the result for $\frac{\pi\langle V_1\rangle}{P_2}$ and $m_{3/2}$ in \cite{Acharya:2008hi}: \ba \frac{\pi\langle V_1\rangle}{P_2} \approx 10.275;\;m_{3/2}\approx 9\times10^5\,{\rm TeV}\,\left(\frac{8\pi^{1/2}\,C_2}{{\cal V}^{3/2}}\right),\ea with $C_2$ an ${\cal O}(1)$ constant, leads to the following upper bound on the light axions from (\ref{massspec}): 
\ba \label{axionmass-max} \hat{m}_{a}^{max} &=& {\cal O}(0.1)\,(m_{3/2}m_{pl})^{1/2}\,e^{-\pi \{V_{j}\}^{min}}\nonumber\\
&=& {\cal O}(1)\,(10^{-8}...1)\,{\rm eV}\,,\ea the precise value of which depends on the stabilized internal volume ${\cal V}$ of the compactification. As a conservative estimate, we take $\{V_{J}\}^{min}=15$, which implies $\hat{m}_{a_K}^{max}\sim 1$ eV. 

The lower bound on light axion masses, although not robustly fixed by theory, can be constrained by phenomenological considerations. Successful gauge coupling unification in the MSSM suggests a unified gauge coupling $\alpha_{GUT}^{-1}\approx 25$, which means that the visible three-cycle volume $V_{vis}\approx 25$. From (\ref{massspec}), this implies the existence of at least one light axion with mass $\hat{m}_{a} \sim 10^{-15}$ eV. In any case, axions lighter than the current Hubble parameter $H_0\approx 10^{-10}\,{\rm yr^{-1}} \sim 10^{-33}$ eV will not have started to oscillate yet and hence are not observable. Thus, for practical purposes, one has: \ba\label{axionmassmin} \hat{m}_{a}^{min} \sim 10^{-33}\,{\rm eV}\ea

We complete this subsection with a technical remark about computing the axion mass eigenstates.
In supergravity, the basis for the moduli and axion fields is in general such that the K\"{a}hler metric is \emph{both} non-diagonal and not canonically normalized (in the sense of giving a canonical kinetic term for scalar fields). Therefore, one has to diagonalize the K\"{a}hler metric by a unitary transformation $U$ as mentioned above and then rescale by $f$ to get a canonically normalized kinetic term. However, this is not the mass eigenstate basis as in this basis the mass matrix is \emph{not} diagonal. Thus, one has to perform a further unitary transformation $\cal{U}$ to find the mass eigenstates. Formally, one can thus relate the axion mass eigenstates $\psi_M;\,M=1,2,...N+1$ to the original axion fields $\hat{t}_L=\{t_i,\theta\};\,i=1,2,..,N$ as:
\ba\label{eigen} \psi_M = {\cal U}^{\dag}_{MK}\,f_K\,U^{\dag}_{KL}\,{\hat t}_L;\;\;\;K,L,M=1,2,..,N+1\,,\ea where ${\cal U}$ diagonalizes the mass-squared matrix : \ba\label{mass-matrix} m_{MN}^2=\frac{1}{f_M f_N}\,U^{\dag}_{MK}V_{KL}U_{LN};\;\;V_{KL}\equiv \frac{\partial^2\,V_{eff}}{\partial \hat{t}_K\partial\hat{t}_L}\,.\ea 

Note that the above analysis only studied non-perturbative effects arising from string instantons and/or hidden sector strong dynamics, and did \emph{not} include the effects of QCD instanton contributions to the scalar potential. We turn to this issue in the next subsection.

\section{The QCD Axion and solution to the Strong CP problem}\label{qcdaxion}

The $M$ theory models under consideration are grand unified theories and, hence,
one would expect the QCD axion to be the
real part of the visible sector gauge kinetic function:
\be
F^{\rm vis}=\sum_{i=1}^NN_i^{\rm vis}z_i\,.
\ee 
However, $\theta_{qcd}\equiv\frac{a_{qcd}}{f_{qcd}}$ at low energies also includes the contribution from phases in the quark mass matrix. In $M$ theory these phases also depend linearly on the $t_i\equiv {\rm Re}(z_i)$ \cite{Acharya:2003gb}, hence taking into account this extra phase shift merely amounts to modifying the coefficients $N_i^{\rm vis}$ to new coefficients $\tilde{N}_i^{\rm vis}$. The QCD axion can then be written as a linear combination of mass eigenstates. Here we have neglected the QCD instanton contributions in computing the mass eigenstates, but we will include them shortly.
\ba\label{bf1}
\theta_{\rm QCD}&\equiv& 2\pi\,{\rm Re}(F^{\rm vis})= \sum_{i=1}^N\tilde{N}_i^{\rm vis}{\tilde t}_i;\;\;{\tilde t}_i\equiv 2\pi t_i\\
{\rm or},\; \theta_{\rm QCD}&=& \sum_{i=1}^N\sum_{K,\,L=1}^{N+1}\tilde{N}_i^{\rm vis}U_{iK}\frac {2\pi}{f_K}{\cal U}_{KL}\psi_L=\sum_{L=1}^{N+1}\frac{\psi_L}{{\tilde f}_L}\,,\nonumber
\ea
using (\ref{eigen}). The effective decay constants ${\tilde f}_L$ are defined by:
\be\label{feff}
\frac 1{{\tilde f}_L}\equiv \sum_{i=1}^N\sum_{K=1}^{N+1}\tilde{N}_i^{\rm vis}U_{iK}\frac {2\pi}{f_K}{\cal U}_{KL}\,.
\ee The effective decay constants ${\tilde f}_L$ depend on the unitary matrices $U$ and ${\cal U}$, which in turn depend on the K\"{a}hler potential. In order to get an idea of the typical size of ${\tilde f}_L$, we do a simple statistical analysis by utilizing features of generic K\"{a}hler potentials consistent with $G_2$ holonomy. The details are given in Appendix \ref{stats}. It turns out that for large classes of generic K\"{a}hler potentials, ${\tilde f}_L$ is close to the standard GUT scale $M_{GUT}$, ${\cal O}(10^{16})$ GeV. This is consistent with expectation, since moduli stablization within this framework generically yields $m_{pl} > M_{11} \gtrsim M_{GUT}$\cite{Acharya:2006ia,Acharya:2008hi}. This also justifies the assumption made in \cite{Arvanitaki:2009fg}. For concreteness we will use ${\tilde f}_L = M_{GUT}=2\times 10^{16}$  GeV in our analysis henecforth. We also use the same value for the individual decay constants $f_L$ for simplicity, which can be justified by the analysis in Appendix \ref{stats}.

We are now ready to include the effects of QCD instantons. The potential from QCD instantons alone is given by (\ref{Vqcd}), expressing it in terms of axion mass eigenstates: 
\ba
V_{qcd}&\sim&\Lambda_{qcd}^4\left(1-\cos\theta_{qcd}\right)\nonumber\\
&\sim&\Lambda_{qcd}^4\left(1-\cos\left(\sum_{L=1}^{N+1}
\frac{\psi_L}{{\tilde f}_L}\right)\right)\,.
\ea
If we were to \emph{disregard} the effects of other contributions to the potential, then the mass-squared matrix for axions has a particularly simple form:
\be\label{nond}
\left(m_{KL}^2\right)_{qcd}\sim\frac{\Lambda_{qcd}^4}{{\tilde f}_K{\tilde f}_L}\,.
\ee
The matrix (\ref{nond}) has one non-zero eigenvalue:
\ba
&&\left(m_{K}^2\right)_{qcd}=0\,,\,\,\,\,\,\forall K=\overline{1,N}\,,\\
&&\left(m_{N+1}^2\right)_{qcd}\sim {\Lambda_{qcd}^4}\sum_{L=1}^{N+1}\frac 1{{\tilde f}_L^2}\,.\nonumber
\ea However, when we do include the effects of the other non-perturbative contributions, the axions will be stabilized as explained in section \ref{spectra}.
These vevs are generally not close to zero; therefore there is a danger that the linear combination 
$\theta_{qcd}$ is not fixed at a value compatible with the experimental
upper bound $|\theta_{qcd}|<10^{-10}$. This, in fact, happens if the masses of \emph{all} 
light eigenstates $\psi_K$ in (\ref{mass-spec}) are heavier than \be \label{mexp} m_{exp}^2\sim 10^{-10}\times\left(m^2_{N+1}\right)_{qcd}\sim (10^{-14})^2\;{\rm eV}^2\,, \ee since then 
the QCD instanton contribution is not strong enough to shift the vevs of $\psi_K$'s away from the values set by the supergravity scalar potential for light axions (\ref{veff}). Interestingly, $m_{exp}$ is close to the mass of the axion for a volume of order $1/\alpha_{GUT}$.

This also implies that, since the QCD axion is represented by a linear combination
of the mass eigenstates, {\it  in order to achieve compatibility with the experimental limit on $\theta_{qcd}$, it would be enough if just one of the mass eigenstates $\{\psi_K\}$
contained inside the linear combination $\theta_{qcd}$ was light compared to $m_{exp}$}. In this case, although the QCD effects will not perturb the vevs of the heavy eigenstates, the vev of the lightest eigenstate contained inside $\theta_{qcd}$ will get readjusted to a new value such that the total linear combination $\theta_{qcd}$ is minimized at $\theta_{qcd}<10^{-10}$. In addition, the mass of this lightest eigenstate will be almost entirely determined by QCD instantons.

From the analysis of the axion spectra in section \ref{spectra} we find that the axion masses are distributed linearly on a logarithmic scale. Thus, by choosing natural values of microscopic parameters in (\ref{mass-spec}), (\ref{Vk}) and (\ref{m32}), we expect to find among the mass eigenstates $\{\psi_K\}$ a number of very light modes with masses smaller than $m_{exp}$, implying that the experimental upper bound on $\theta_{qcd}$ can be easily satisfied. In particular, taking into account the effects of the QCD instantons in the full mass-squared matrix will drastically modify the lightest eigenvalue, which now gets a mass $m_a^{qcd}\sim \frac{\Lambda_{qcd}^2}{f_a^{qcd}}$,  and will modify all eigenstates with mass less than $m_{exp}$. The eigenstate with the mass $m_a^{qcd}$ will then be \emph{uniquely} identified as the QCD axion for all practical purposes.

In Appendix \ref{toy}, we consider a toy example in detail in which we compute the eigenvalues and eigenstates of the axions numerically. The numerical results are completely consistent with the general results described above.

\section{Cosmological Evolution and Relic Abundance}\label{cosmo}

We now study the cosmological evolution of the $N$ axions and compute their relic abundance. However, doing so requires a knowledge of the cosmological history as well as a knowledge of the mass spectrum of moduli vis-a-vis the Hubble parameter during inflation $H_I$. Although $H_I$ is not determined from observations at present, it is possible to get an idea about its magnitude. It is known that the slow-roll parameter for simple models of inflation - $\epsilon
\equiv m_{pl}^2\left(\frac{V'}{V}\right)^2$, where $V$ is the slow-roll inflaton potential and $V'$ is the derivative of $V$ with respect to the inflaton field, can be written in terms of $H_I$ as: \ba\label{epsilon} \epsilon \approx 10^{10}\,\left(\frac{H_I}{m_{pl}}\right)^2\ea using the experimental value of primordial density perturbations $\delta \rho/\rho \sim 10^{-5}$. This implies that $H_I \lesssim 10^{-6}\,m_{pl}$ using the 
fact that $\epsilon \lesssim 10^{-2}$ is required for $\sim$60 e-foldings of inflation to solve the flatness and horizon problems. This is the standard fine-tuning required in slow-roll inflation models. A smaller value of $H_I$ than above will make $\epsilon$ even smaller, implying a larger fine-tuning for $\epsilon$ than is required for inflation. In this paper, we will take an unbiased approach and discuss both cases with a large $H_I (> M_{moduli}$) 
and with a small $H_I (< M_{moduli})$, since the cosmological consequences are qualitatively different. We discuss cosmological observables, which could distinguish between the two situations. 
%The moduli mass spectrum and their effects on cosmological evolution in the framework considered here have been studied in detail in \cite{Acharya:2008bk}. It turns out that all but one moduli (including the meson field $\phi_1$) in (\ref{W}) are stabilized with masses of ${\cal O}(m_{3/2})$ while the remaining modulus and the axion combination $(b_1-b_2)\vec{N_1}\cdot\vec{t}+a\theta$ in (\ref{co}) are stabilized with a mass ${\cal O}(100\,m_{3/2})$. 

\subsection{Relic Abundance in Non-Thermal Cosmology}\label{relic-nonstd}

We first assume that the Hubble parameter during inflation is large, i.e. $H_I > M_{moduli}={\cal O}(m_{3/2})$, with $m_{3/2} \sim$ 10 TeV required for low-scale supersymmetry with gravity mediation, which is natural within this framework \cite{Acharya:2006ia, Acharya:2008hi}. For notational convenience, this cosmological scenario will be dubbed ``non-thermal" cosmology. In this case, there generically exists at least one set of moduli $X_{light}$ that are lighter than $H_{I}$, which in turn will generically be displaced from their late-time minima during inflation. After the end of inflation, the Hubble parameter keeps decreasing and will eventually become comparable to the masses of these moduli, at which time they will start coherent oscillations. Since these oscillations scale like matter, they will quickly dominate the energy density of the Universe. The lightest of these moduli will start oscillating last. Since the moduli are very weakly coupled to the visible sector, they will decay long after all the moduli have started oscillating. The requirement that the decay of the lightest modulus $X_{0}$ gives rise to a reheat temperature greater than a few MeV for successful BBN puts a lower bound on the mass of $X_{0}$. Within gravity mediation, one finds $m_{X_0} \sim m_{3/2} \gtrsim {\cal O}(10)$ TeV, the precise value depending on model-dependent details. It was shown in \cite{Acharya:2008bk} that both the moduli and gravitino problems can be naturally solved within this framework.

In order to compute the relic abundance of axions in such a framework, it is important to consider two era's: before and after the moduli have decayed. During the first period the Universe is moduli dominated.
This period then gives way to a radiation-dominated era after the decay of the lightest modulus. 
It is this latter period in which BBN and later cosmological events such as recombination, matter-radiation equality, and growth of structure take place. The computation of the relic-abundance of axions in these two regimes is very different, which we now study. 

\subsubsection{Radiation-dominated Era}\label{rad}

If the mass of a given axion is such that it starts oscillating \emph{after} the lightest modulus decays, i.e when $\hat{m}_a < \Gamma_{X_0}={\cal O}(1)\frac{m_{X_0}^3}{m_{pl}^2}$, the standard computation of the relic abundance is applicable. This gives the following boundary value: \ba\label{boundary} \bar{m} = {\cal O}(1-10)\times10^{-15}\;{\rm eV}\ea 

For axions with masses below $\bar{m}$, the axion relic abundance can be computed as: \ba \label{thermal}\Omega_{a_k}\,h^2 = 0.06\left(\frac{\hat{f}_{a_k}}{2\times 10^{16}\,{\rm GeV}}\right)^2\left(\frac{m_{a_k}}{10^{-20}\,{\rm eV}}\right)^{1/2}\,\langle\theta_{I_k}^2\rangle\,\chi\ea where $\chi$ is an ${\cal O}(1)$ fudge factor to take into account possible effects from anharmonicity, etc. The masses of these axions are due to string/membrane instantons and are much smaller than that of the QCD axion ($\approx 3\times 10^{-10}$ eV), hence they are not expected to receive corrections from finite temperature effects.  It is important to note the dependence of the relic abundance on $\hat{f}_{a_k}$ and $\hat{m}_{a_k}$. The relic abundance \emph{increases} by increasing the mass and the decay constant. The upper bound on the relic abundance therefore implies that for $\hat{f}_{a_k} \approx 2\times 10^{16}$ GeV and $\langle\theta_{I_k}^2\rangle = {\cal O}(1)$, there is an upper bound on the mass of the axion: \ba \label{thermal-2}m^{(std)}_{relic} = {\cal O}(1)\times 10^{-20}\;{\rm eV}\ea

Thus, from (\ref{boundary}) and (\ref{thermal-2}), the misalignment angle $\langle\theta_{I_k}^2\rangle$ has to be fine-tuned for axion masses between $10^{-20}\lesssim \hat{m}_{a_k} \lesssim 10^{-14}$ eV.

\subsubsection{Moduli-dominated Era}\label{mod}

For axions with masses $\hat{m}_{a_k} \gtrsim 10^{-14}$ eV, the relic abundance is determined by a different computation since the Universe is moduli-dominated. The abundance can be readily computed \cite{Fox:2004kb}: \ba\label{mod-relic} \Omega_{a_k}\,h^2 &=& {\cal O}(1)\;\left(\frac{T_{RH}^{X_0}\hat{f}_{a_k}^2}{m_{pl}^2\,(3.6\,{\rm eV})}\right)\,\langle\theta_{I_k}^2\rangle\,\chi\\&=&{\cal O}(10)\,\left(\frac{\hat{f}_{a_k}}{2\times10^{16}{\rm GeV}}\right)^2\left(\frac{T_{RH}^{X_0}}{1\,{\rm MeV}}\right)\langle\theta_{I_k}^2\rangle\,\chi\nonumber\ea Note that the relic abundance is completely \emph{independent} of the mass of the axion; apart from $\langle\theta_{I_k}^2\rangle$ it only depends on the (effective) decay constant, which is approximately $M_{GUT}\equiv 2\times 10^{16}$ GeV within our framework, and the final reheat temperature $T_{RH}^{X_0}$,  which is more or less around 1 MeV. Thus, \emph{the relic abundance for \emph{all} axions with masses $\hat{m}_{a_k} \gtrsim 10^{-14}$ eV is independent of their masses and only depends on the misalignment angle $\langle\theta_{I_k}^2\rangle$}. Noteice that, within our framework, the mass of the QCD axion (the mass eigenstate that dominantly gets its mass from QCD instantons) automatically lies in this region, $m_a^{qcd}\approx 3\times 10^{-10}\,{\rm eV} > 10^{-14}\,{\rm eV}$. Hence its relic abundance is just determined by $\langle\theta_{I}^2\rangle_{a^{qcd}}$. In order to be consistent with the WMAP upper bound ($\Omega_{DM}h^2 \leq 0.11$), this requires:
\ba \langle\theta_I^2\rangle_{a^{qcd}} \lesssim 10^{-2}\,,\ea which implies a modest fine-tuning in the intial conditions. We will discuss this in detail in section \ref{Cons}.

\subsection{Relic Abundance in `Thermal' Cosmology}\label{relic-std}

The cosmological framework studied above can be contrasted with one in which $H_I < M_{moduli}={\cal O}(10)$ TeV,  which will be termed as "thermal" cosmology for notational purposes,. This corresponds to a situation with a much smaller $H_I$ and hence a much smaller reheat temperature after inflation $T_{RH}^I$. Now, a radiation-dominated phase follows after reheating (from inflation) at $T_{RH}^I$ and continues until matter-radiation equality at $T_{EQ}$. In this case the relic abundance of \emph{all} axions is computed as it was done in section \ref{rad}, giving rise to the same equation: \ba \label{thermal3}\Omega_{a_k}\,h^2 = 0.06\left(\frac{\hat{f}_{a_k}}{2\times 10^{16}\,{\rm GeV}}\right)^2\,\left(\frac{m_{a_k}}{10^{-20}\,{\rm eV}}\right)^{1/2}\langle\theta_{I_k}^2\rangle\,\chi\ea For the QCD axion, finite-temperature mass effects for the QCD axion with $\hat{m}_{a_{QCD}}\approx 3\times 10^{-10}$ eV will modify the dependence on $\{\hat{f}_{a_k},\hat{m}_{a_k}\}$ compared to that in (\ref{thermal}). Thus, for the QCD axion one finds \cite{Fox:2004kb}:
\ba
\Omega_{a_{QCD}}h^2 \approx 4.5\times10^4\left(\frac{\hat{f}_{a_{QCD}}}{2\times 10^{16}{\rm GeV}}\right)^{7/6}\langle \theta^2\rangle \chi \nonumber
\ea Note that (\ref{thermal3}) means that the relic abundance for each axion increases with the mass of the axion in contrast to that within non-thermal cosmology, giving rise to an upper bound for the axion mass $\hat{m}^{(std)}_{relic} \sim 10^{-20}$ eV, above which $\langle\theta_{I_k}^2\rangle$ has to be tuned.

\section{Other Cosmological Constraints}\label{Cons}

In addition to the relic abundance constraint, there are other constraints arising from the presence of light axions. The first is the presence of an isocurvature component of temperature fluctuations ($\alpha_a$) and the second is the presence of a non-Gaussian component of temperature fluctuations, both arising from fluctuations of the axions during inflation. These constraints have been studied earlier in various contexts. For example, \cite{Fox:2004kb} has studied the constraints for the QCD axion with large $\hat{f}_{a}^{qcd}$ for both thermal and non-thermal cosmological histories. A recent paper \cite{Hamann:2009yf} studies the consequences for iscourvature fluctuations within the above assumptions for standard cosmology. \cite{Hertzberg:2008wr} has studied the constraints for the QCD axion with thermal cosmology as a function of the decay constant $\hat{f}_{a}^{qcd}$. \cite{Visinelli:2009kt} has studied constraints for the QCD axion for both standard and non-standard cosmological histories assuming that the QCD axion comprises all of DM. \cite{Mack:2009hs} has studied constraints arising from multiple axions, but with different assumptions than the one considered in this paper. In our analysis, we have fixed $\hat{f}_{a_k}$ to be the GUT scale as it is the natural scale in the framework considered. We then generalize the analysis of \cite{Fox:2004kb} for the QCD axion to a situation with many axions roughly distributed evenly on a logarithmic scale.  

The observables above depend on the axion relic abundances ($\Omega_{a}h^2$) and the Hubble parameter during inflation ($H_{I}$). Furthermore, the gravity wave contributions to temperature fluctuations also depend on $H_{I}$. It turns out that the bound on a non-Gaussian component in the CMB does not give rise to any additional constraints on the parameter space, so we consider the following:
\ba\label{constraints}
\Omega_a h^2 &\equiv& \sum_{k=1}^N \Omega_{a_k} h^2 \leq 0.11\\ 
\alpha_a &\equiv& \sum_{k=1}^N\,\frac{8}{25}\,(\frac{(\Omega_{a_k}/\Omega_m)^2}{\langle (\delta T/T)^2_{tot}\rangle})\;\sigma_{\theta_k}^2\,(2\theta_{I_k}^2+\sigma_{\theta_k}^2) \leq 0.072\nonumber\\
Q_t &\equiv& \frac{H_{I}}{5\pi\,m_{pl}}\leq 9.3\times 10^{-6}\nonumber
\ea
where we have used the latest bounds from WMAP5 \cite{Komatsu:2008hk}. Note that the quantity $\langle \theta_{I_k}^2 \rangle$ appearing in the expression for the relic abundance of axions is given by: \ba \langle \theta_{I_k}^2 \rangle &\equiv& \theta_{I_k}^2+\sigma_{\theta_k}^2\nonumber\\ {\rm where}\;\;\sigma_{\theta_k}&\equiv& \frac{H_{I}}{2\pi\,\hat{f}_{a_k}}\ea The observed upper bounds on the relic abundance ($\Omega_a h^2$), tensor modes ($Q_t$) and isocurvature fluctuations ($\alpha_a$) provide a constraint on the $2N+1$ microscopic parameter space - $\{\hat{m}_{a_k},\theta_{I_k},H_{I};\,k=\overline{1,N}\}$ in general, where $\theta_{I_k}$ is the initial {\it mean} misalignment angle of the axion $a_k$. More precisely, a given spectrum of axions $\{\hat{m}_{a_k}\}$ imposes constraints on the parameters $\{\theta_{I_k},H_{I}\}$. 

From our understanding of the mass spectra of axions in section \ref{spectra}, it is possible to dramatically reduce the number of parameters, as follows. Using (\ref{axionmass-max}) and (\ref{axionmassmin}), 
\ba  
10^{-33}\,{\rm eV}&\lesssim& \hat{m}_{a_k}\lesssim 1\,{\rm eV};\;\;\;\;k=\overline{1,N} \nonumber\\ 39 &\gtrsim& V_j \gtrsim 15;\;\;\;\;\;\;\;j=\overline{3,N+2}
\ea 
Note that for non-thermal cosmology, the boundary between the moduli-dominated and radiation-dominated 
regimes given by $\bar{m}={\cal O}(1-10)\times 10^{-15}$ eV in (\ref{boundary}), corresponds to $\bar{V}_J \approx 25$. Now, since $\hat{m}_{a_k} \propto 
e^{-\pi V_{K+2}}$ and $V_{K+2}$ varies by ${\cal O}(1)$ for different three-cycles in the compactification, we assume that there are ${\cal O}(1)$ axions in each e-folding between $\sim 1$ eV and $\sim 10^{-33}$ eV, corresponding to $15\lesssim V_{K+2}\lesssim 39$. This is expected to be true provided the total number of axions $N$ is sufficiently large. For simplicity we also assume that the initial mean misalignment angles of all axions are roughly equal: $\theta_{I_k}\approx\theta_{I_0},\,k=\overline{1,N}$. 

With these assumptions, the number of microscopic parameters is reduced to \emph{four} for non-thermal cosmological evolution - $\{\theta_{I_0},H_I,N_1,N_2\}$, where $N_1$ and $N_2$ are the number of axions with masses corresponding to $V_{K+2}$ in the ranges $15 \lesssim V_{K+2} \lesssim \bar{V}_{K+2}(\approx 25)$ (moduli-dominated regime) and $25 \lesssim V_{K+2} \lesssim 39$ (radiation-dominated regime) respectively. For thermal cosmology, the number of parameters is just \emph{three} - $\{\theta_{I_0},H_I,N_{std}\}$, where $N_{std}$ is the number axions with masses corresponding to $V_{k+2}$ in the range $15 \lesssim V_{k+2} \lesssim 39$. In particular, with \emph{one} axion in each e-folding between $\sim 1$ eV and $\sim 10^{-33}$ eV, $N_1 \approx 10,\,N_2 \approx 14$, and $N_{std}=24$.

\begin{figure}[h!]
%\begin{tabular}{c}
%\leavevmode \epsfxsize 9 cm \epsfbox{mass-Vk.eps}
\resizebox{8cm}{!}{\includegraphics{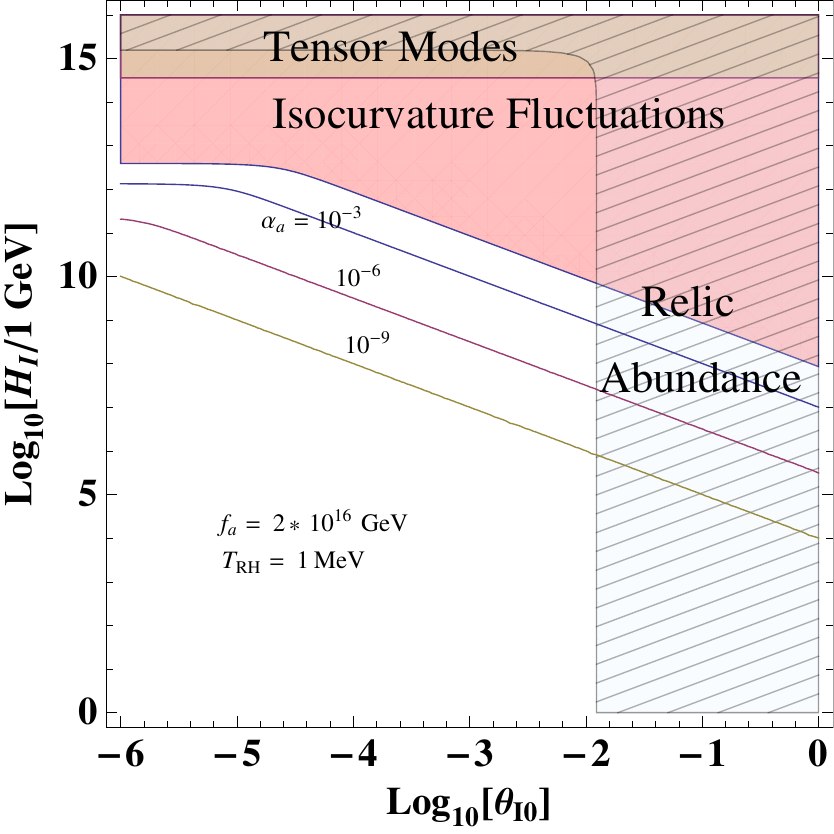}}
%\resizebox{8cm}{!}{\includegraphics{axion-constraints-std.eps}}
%\end{tabular}
\caption{Allowed microscopic parameter space (unshaded region) in the $\{\theta_{I_0},H_I\}$ plane for $N_1=10,\,N_2=14$ with a ``non-thermal", moduli dominated cosmological history ($H_I > M_{moduli}$) after imposing the current bounds on tensor modes, isocurvature fluctuations and the overall relic abundance. Contours for three allowed values of the isocurvature fluctuations $\alpha_a$ are also plotted.}\label{axion-constraints}
\end{figure}
Figure \ref{axion-constraints} shows the effect of the cosmological constraints on the two dimensional parameter space $\{\theta_{I_0},H_I\}$ for non-thermal cosmology with $N_1=10$ and $N_2=14$. The range of $H_{I}$ is chosen to be between $10^5$ and $10^{16}$
GeV. The lower bound is so chosen because the non-thermal cosmology occurs when $H_{I} > M_{moduli}$. It can be seen from the figure that the bound on tensor modes excludes $H_I \gtrsim 3\times 10^{14}$ GeV, while the combination of the isocurvature and relic density bounds imply $H_I \lesssim 10^{10}$ GeV and $\theta_{I_0} \lesssim 10^{-2}$. Thus, a modest fine-tuning (percent level) in $\theta_{I_0}$ is required. 

It is instructive to compare the results obtained in Figure $\ref{axion-constraints}$ with those in the thermal cosmology, i.e. in which $H_I < M_{moduli}$. As explained earlier, in this case the moduli are not displaced from their late-time minima during inflation, 
and a standard thermal history with a radiation-dominated phase follows after the end of inflation. With $M_{moduli}\gtrsim m_{3/2} \gtrsim$ 10 TeV as in our framework, 
$H_I$ is bounded from above by approximately $10^5$ GeV. Assuming that the reheating process after the end of inflation is efficient, this implies that the reheat temperature after inflation $T_{RH}^I \lesssim 2\times 10^{11}$ GeV. Then, the thermal abundances of the gravitino and axinos/modulinos (also with mass $\approx m_{3/2}$ in our framework \cite{Acharya:2008bk}) are such that they do not overproduce LSPs by their decays \cite{Kawasaki:2008qe,Kawasaki:2007mk}. Moreover, since $m_{3/2} \gtrsim 10$ TeV, they decay before the onset of BBN.

\begin{figure}[h!]
%\begin{tabular}{c}
%\leavevmode \epsfxsize 9 cm \epsfbox{mass-Vk.eps}
%\resizebox{8cm}{!}{\includegraphics{axion-constraints.eps}}
\resizebox{8cm}{!}{\includegraphics{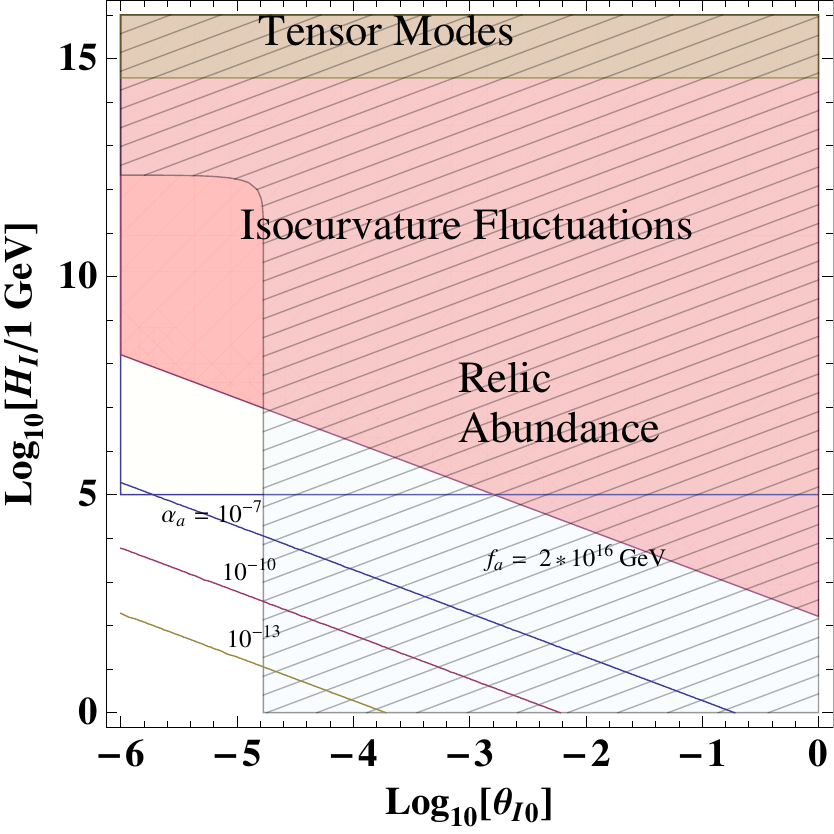}}
%\end{tabular}
\caption{Allowed microscopic parameter space (unshaded region) in the $\{\theta_{I_0},H_I\}$ plane for $N_{std}=24$ with a ``thermal" cosmological history ($H_I < M_{moduli}$) after imposing the current bounds on tensor modes, isocurvature fluctuations and the overall relic abundance. Contours for three allowed values of the isocurvature fluctuations $\alpha_a$ are also plotted.}\label{axion-constraints-std}
\end{figure}

Figure \ref{axion-constraints-std} shows the constraints on the microscopic parameters $\{\theta_{I_0},H_I\}$ taking into account the above effects with the same mass distribution of axions, i.e. with one axion in each e-folding between $\sim 1$ eV and $\sim 10^{-33}$ eV. Then, the number of axions $N_{std}$ in the entire mass range correspond to $15 \lesssim V_k \lesssim 39$, implying $N_{std}=24$.
 The relic abundance bound requires $\theta_{I_0} \lesssim 10^{-5}$, while the isocurvature bounds are automatically satisfied by imposing the relic abundance bound and the requirement $H_I < M_{moduli}$. This further implies that isocurvature fluctuations can only be observed within thermal cosmology with $\alpha_a \lesssim 10^{-7}$.

The differences between Figures \ref{axion-constraints} and \ref{axion-constraints-std} are clear. In Figure \ref{axion-constraints-std}, the relic density bound implies a fine-tuning in $\theta_{I_0}$, which is a little less than three orders of magnitude worse compared to that for Figure \ref{axion-constraints}. This is because the relic abundance is a monotonically increasing function of the axion mass in Figure \ref{axion-constraints-std} for {\it all} axions, while it is {\it independent} of the axion mass (for $\hat{m}_{a_k} \gtrsim 10^{-14}$ eV) in Figure \ref{axion-constraints}. This is crucial because the largest contribution to the relic abundance in Figure \ref{axion-constraints-std} comes from these heavier axions. From (\ref{epsilon}), the requirement $H_I < M_{moduli}$ suggests that the fine-tuning in $\epsilon$ (at least for simple inflationary models) is much worse than that in Figure \ref{axion-constraints}. Our results are consistent with the earlier results of reduced fine-tuning for a single (QCD) axion coming from entropy production due to late decay of scalar condensates \cite{Fox:2004kb}, \cite{latedecay}, and generalizes those results to the case with a plethora of axions. Finally, the magnitude of allowed isocurvature fluctuations is about five orders of magnitude smaller for thermal cosmology compared to that for non-thermal cosmology. 

In the above, we have assumed the existence of one axion in each e-folding between $\sim 1$ eV and $\sim 10^{-33}$ eV. However, it could happen that $V_K$ for different three-cycles in the internal manifold scans less finely (but still varies by ${\cal O}(1)$), giving rise to say, one axion in every \emph{ten} e-foldings. In addition, as mentioned below 
(\ref{axionmass-max}), depending upon the details of the compactification ${\cal V}$ could be stabilized at values close to its upper bound, leading to a larger lower bound for $V_k$ and hence a smaller $\hat{m}_{a_k}^{max}$.
These effects will make $\{N_1,N_2\}$ ($N_{std}$) smaller than that assumed in Figure 1 (Figure 2), and could help in relaxing the constraints on $\{\theta_{I_0},H_I\}$. 

\begin{figure}[h!]
\centering
\begin{tabular}{cc}
\epsfig{file=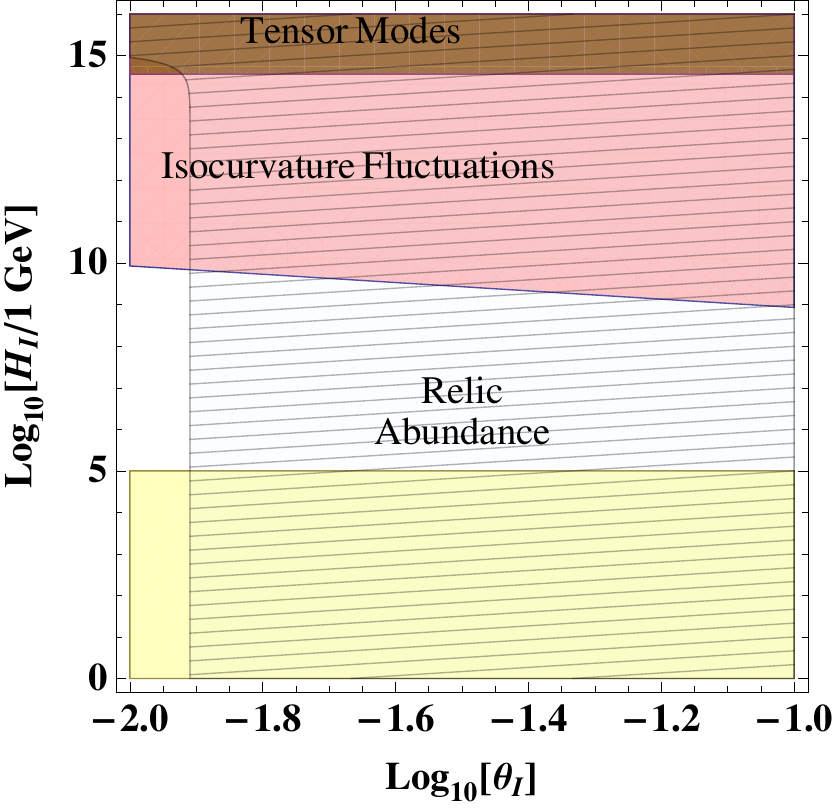,width=0.5\linewidth,height=0.28\textwidth,clip=} &
\epsfig{file=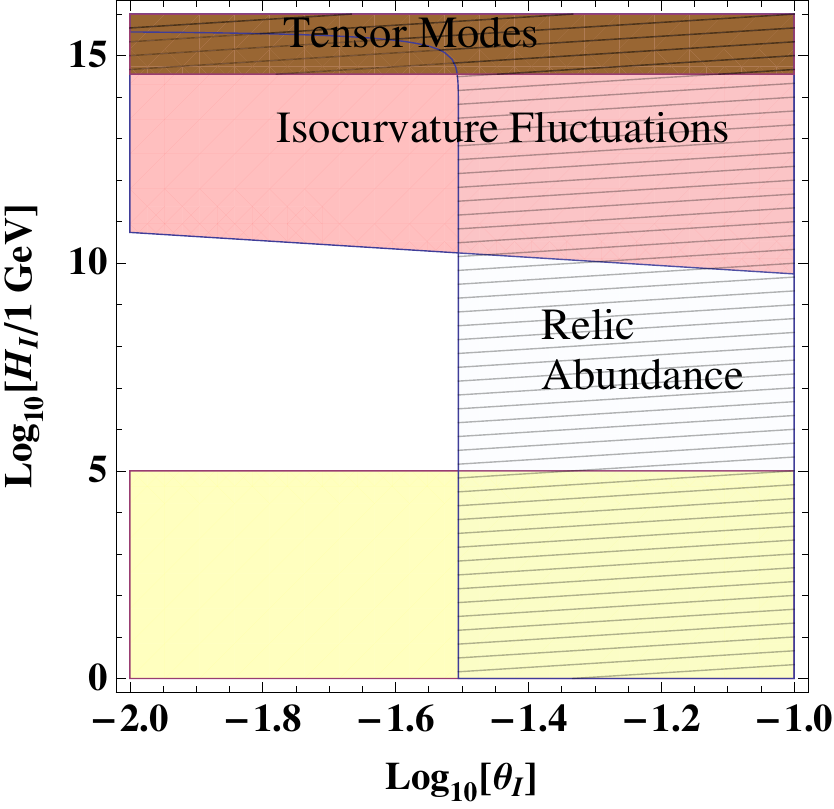,width=0.5\linewidth,height=0.28\textwidth,clip=}
\end{tabular}
\caption{Effect on allowed values of microscopic parameters $\{\theta_{I_0},H_I\}$ by decreasing  
$\{N_1,N_2\}$, for non-thermal cosmological history ($H_I > M_{moduli}$).
{\bf Left}: $N_1=10,\,N_2=14$; {\bf Right}: $N_1=1,N_2=2$.}\label{axconst}
\end{figure}
\begin{figure}[h!]
\centering
\begin{tabular}{cc}
\epsfig{file=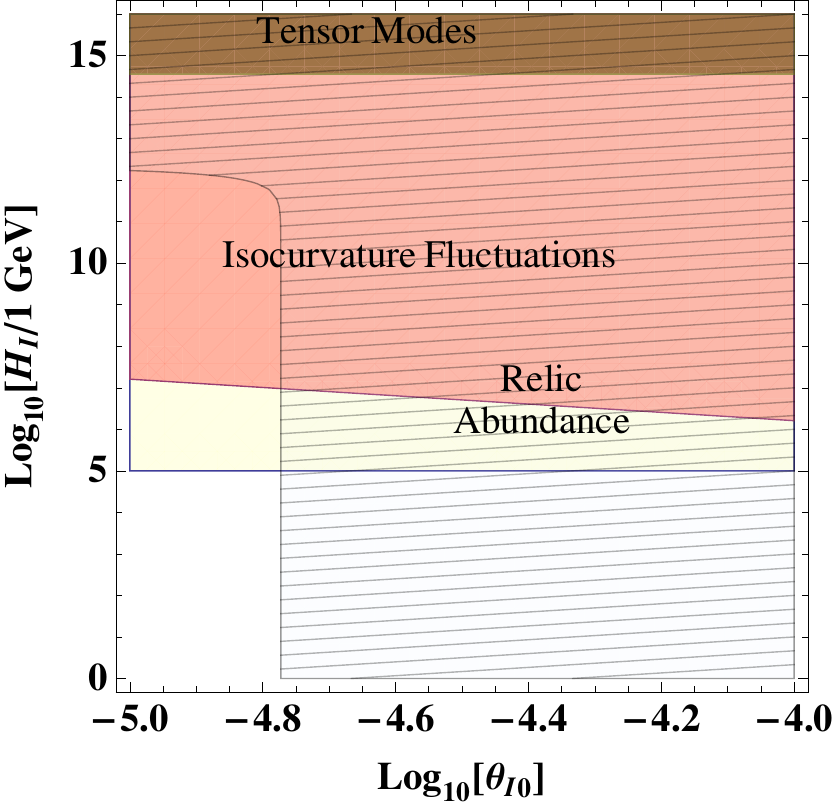,width=0.5\linewidth,height=0.28\textwidth,clip=} &
\epsfig{file=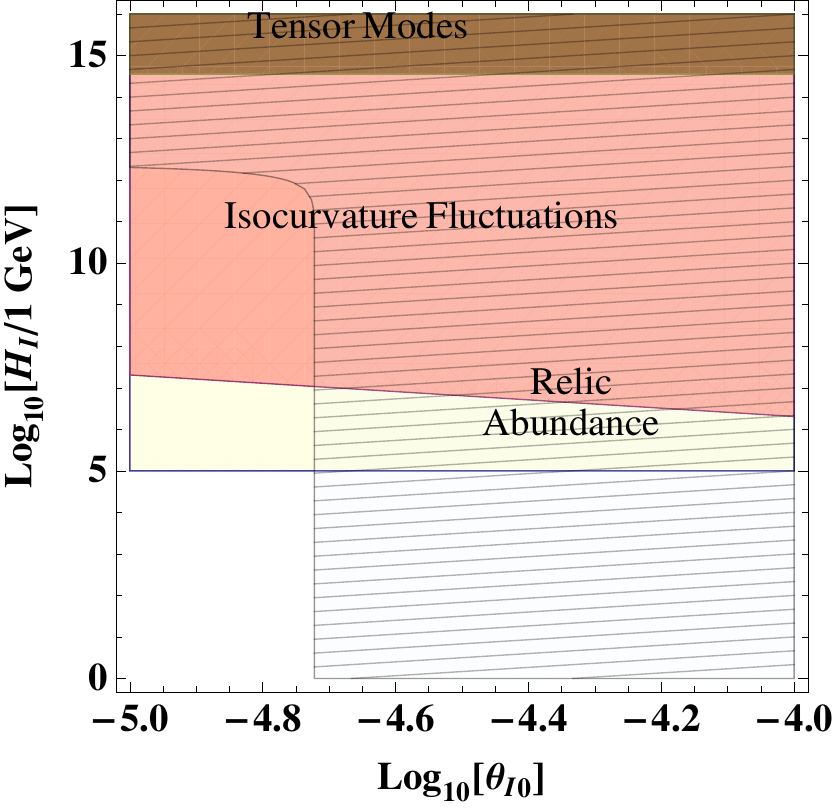,width=0.5\linewidth,height=0.28\textwidth,clip=}
\end{tabular}
\caption{Effect on allowed values of microscopic parameters $\{\theta_{I_0},H_I\}$ by decreasing  
$N_{std}$, for thermal cosmological history ($H_I < M_{moduli}$).
{\bf Left}: $N_{std}=24$; {\bf Right}: $N_{std}=3$.}\label{axconst-std}
\end{figure}

In Figures \ref{axconst} and \ref{axconst-std}, we show the effects of decreasing $\{N_1,N_2\}$ from $\{10,14\}$ to $\{1,2\}$, and $N_{std}$ from 24 to 3, respectively. Within non-thermal cosmology this has the effect of relaxing the constraints on $\{\theta_{I_0},H_I\}$, as seen from Figure \ref{axconst}. However, the constraints for thermal cosmology shown in Figure \ref{axconst-std} are essentially unchanged.

The above analysis shows that a String Axiverse with a large
Hubble parameter during inflation $H_I > M_{moduli}$ is much less constrained than in the alternative case.

\subsection{Consequences}\label{consequences}

It is important to understand and appreciate observables which could falsify the entire approach
as well as distinguish among the two cosmological histories. From the third equation in (\ref{constraints}) and from Figures \ref{axion-constraints} and \ref{axion-constraints-std}, it can be seen that the observation of tensor modes in the future by PLANCK requires a reasonably large $H_I \gtrsim 3\times 10^{12}$ GeV, which is conclusively ruled out within our approach for both cosmological histories. Hence, \emph{an observation of tensor modes in the near future in general, and by PLANCK in particular, will rule out the String Axiverse}. This is a similar conclusion to \cite{Fox:2004kb}, but now it applies to the entire Axiverse.
Note also that isocurvature fluctuations can effectively discriminate between the ``thermal" and ``non-thermal" cosmological histories considered in this paper. As can be seen from Figures \ref{axion-constraints} and \ref{axion-constraints-std}, non-thermal cosmology can give rise to isocurvature fluctuations just below the observed bound $\alpha_a^{non-std}\lesssim 0.072$, while thermal cosmology predicts isocurvature fluctuations which are vastly suppressed, $\alpha_a^{std} \lesssim 3\times 10^{-7}$. \emph{Therefore, an observation of isocurvature fluctuations in the near future will rule out a String Axiverse with thermal cosmology }. On the other hand, although a lack of observation of isocurvature fluctuations in the near future will not rule out non-thermal cosmology within the approach, it will disfavor it.

\subsection{Constraints from Production of Light Axions from Other Sources}\label{axionprod}

The relic abundance of light axions in the previous sections has been computed assuming that the axions act as coherent classical fields with zero momentum. However, in general there are two other contributions to the axion relic abundance:
\begin{itemize}
\item Light axions could be produced during interactions among particles in the thermal plasma created during reheating after inflation. 
\item Light axions could be produced directly from the decay of moduli (scalar fields) with masses $\sim m_{3/2}$. 
\end{itemize}

The production of light axions in the thermal plasma has been studied in \cite{Kolb:Turner} for ``thermal" cosmology. It turns out that axions with $\hat{f}_{a_k}\sim 2\times 10^{16}$ GeV  interact so weakly with the thermal plasma that a thermal population of axions never results.

Within ``non-thermal" cosmology, these thermal axions, even if present, will be vastly diluted by the decay of moduli. So they are completely irrelevant. However, in this case light axions can be produced from the decay of coherently oscillating scalar fields displaced during inflation. The axions thus produced contribute to the total number of \emph{effective} neutrino species $N^{eff}_{\nu}=N_{\nu}+\Delta N_{\nu}$ for which there is an upper bound from BBN due to $^4He$ overproduction as well as from CMB measurements ($\Delta N_{\nu} \lesssim 1$) \cite{Kawasaki:2007mk, Komatsu:2008hk}. Hence, it is important to check if these bounds can be satisfied within our framework. 

The relevant scalar field to consider is the one that is the lightest and decays \emph{last}, since the axions possibly produced from the decay of heavier moduli will be vastly diluted by entropy production of the lighter ones. As noted earlier, within the scheme of moduli stabilization considered here, $N-1$ moduli are stabilized with masses $\approx (1-2)\,m_{3/2}$ \cite{Acharya:2008bk}. In cases with matter axions present in the subdominant condensates, it turns out that the matter axions are also stabilized with masses $\approx m_{3/2}$. If it happens that the lightest scalar field $X_0$ is an axion rather than a geometric modulus, then it cannot decay to two light axions because of the derivative nature of the axion coupling. In this case, there is negligible production of ultra-relativistic light axions. On the other hand, if $X_0$ is a geometric modulus, then a tree-level decay of this field to two light axions is possible via the operator : $C_k\,X_0\,\partial_{\mu}\hat{a}_k\partial^{\mu}\hat{a}_k$ for some model-dependent coefficient $C_k$. Since the decaying scalar field dominates the energy density of the Universe, the yield $Y_{a} = \Delta N_{\nu}$ of light axions is given by: \ba 
Y_{a} = B_{a}\,Y_{X_0} = B_{a}\,\frac{3}{4}\left(\frac{T_{RH}}{M_{X_0}}\right)\ll 1 \ea where $B_a$ is the branching ratio of $X_0$ to all axions - $X_0 \rightarrow \hat{a}_k\,\hat{a}_k;\, k=\overline{1,N}$. Here $T_{RH}$ is the reheat temperature from the decay of the lightest modulus $X_0$, which is ${\cal O}(1)$ MeV for $M_{X_0} \gtrsim 10$ TeV. The bound is thus easily satisfied.

\section{Observables}\label{observables}

We finish the paper by returning to the observable consequences of the String Axiverse that were discussed in \cite{Arvanitaki:2009fg}.
The basic prediction for the Axiverse spectrum we find is that the masses are distributed roughly evenly on a logarithmic scale between 
$\sim 1$ eV and $\sim 10^{-33}$ eV. 

As explained in \cite{Arvanitaki:2009fg}, the mass spectrum of axions can be divided into four windows as far as observable effects are concerned. If there are axions in the window $10^{-33}\,{\rm eV} \lesssim \hat{m}_{a_k} \lesssim 4\times 10^{-28}\,{\rm eV}$, which couple appreciably to $\vec{E}\cdot \vec{B}$, they could cause a rotation of the polarization of the CMB. Could such axions arise within the framework considered here? Unfortunately not. The $M$ theory models have Standard Grand Unification and hence the axion that predominantly couples to  $\vec{E}\cdot \vec{B}$ is the QCD axion, whose mass is too large ($\approx 3\times 10^{-10}$ eV). Since the QCD axion is a linear combination of all axion mass eigenstates one might obtain couplings of lighter axions to $\vec{E}\cdot \vec{B}$, however these are suppressed by
${\cal O}(\frac{\hat{m}_{a}^{light}}{\hat{m}_{a}^{heavy}})^2$, as
confirmed in the toy example considered in Appendix \ref{toy}. 
Hence, one does not expect a rotation of the polarization of the CMB.

Axions in the mass window $10^{-28}\,{\rm eV} \lesssim \hat{m}_{a_k}< 3\times10^{-18}$ eV can give rise to step-like features in the matter power spectrum at small scales. The physics underlying this effect is the following: for very light scalar fields, there is a mass-dependent pressure term in the Euler equation governing the density fluctuations, which gives rise 
to a modified Jeans length, below which density fluctuations do not grow. It is very 
similar to the suppression from free-streaming due to light neutrinos. Such axions can naturally arise within our framework.

Axions in the mass window $10^{-18}\lesssim \hat{m}_{a_k}\lesssim 10^{-10}$ eV can form exponentially growing bound states with rotating black holes and hence significantly affect their dynamics by graviton emission that carries aways the black hole's angular momentum \cite{Arvanitaki:2009fg}. This causes a spin-down of the black hole, and occurs for black holes lighter than $\sim 10^7\,M_{solar}$, (corresponding to axion masses heavier than $\sim 10^{-18}$ eV), resulting in gaps in the mass spectrum of rapidly rotating black holes. A recent paper \cite{Arvanitaki:2010sy} studies other interesting signals such as gravitational waves and gamma rays emanating from this axion-black hole bound state under particular conditions. These can supposedly probe axions with masses upto $\sim 10^{-10}$ eV, the mass of the QCD axion. Since the effect only depends on the mass of the axion, the signal is the same for both cosmological histories considered.

Finally, if axions with masses in the range $10^{-10}\lesssim \hat{m}_{a_k}\lesssim 1$ eV have an appreciable coupling to $\vec{E}\cdot\vec{B}$, they may be detected by their decays to photons in our galaxy or by spectroscopic effects in compact astrophysical environments of magnetars, pulsars and quasars arising from photon-axion conversion in the strong magnetic field present inside these objects \cite{Raffelt}. The latter may be crucial in probing large axion decay constants $\hat{f}_{a_k} \gg 10^9$ GeV \cite{Chelouche:2008be}. Do we expect such axions within our framework? As explained earlier, before turning on QCD instanton effects the axionic partner of the modulus measuring the QCD gauge coupling is an ${\cal O}(1)$ linear combination of all axion mass eigenstates. From the argument in the paragraphs above, we saw that after turning on QCD effects eigenstates, which are much lighter than $m_a^{qcd}\sim 10^{-10}$ eV do not couple appreciably to $\vec{E}\cdot\vec{B}$. However, eigenstates that are much heavier than $10^{-10}$ eV can still couple appreciably to $\vec{E}\cdot\vec{B}$ since they are hardly affected by QCD instantons. This can also be checked from the explicit example studied in Appendix \ref{toy}. Thus, we expect a few axions in the range $10^{-10}$ eV $< \hat{m}_{a_k} \lesssim 1$ eV coupling appreicably to $\vec{E}\cdot\vec{B}$. The feasibility of such signals, however, depends on the strength of axion-photon coupling ($\sim \hat{f}_a^{-1}$) and the reduction of backgrounds.

It is interesting to note that if the {\it lightest} axion in the Axiverse has a mass corresponding to a GUT instanton, i.e. $m_{a_k}> 10^{-15}$ eV for all axions and if additionally the axion decay constants  $f_a$ are all of order $10^{14}$ GeV, then there is no fine-tuning required at all to satisfy the cosmological constraints in the non-thermal, moduli dominated cosmology. In this case, the fundamental scale is also of order $10^{14}$ GeV. Although not fine tuned, this case may be difficult to reconcile with both unification and bounds on the proton lifetime within standard GUTs. It has been argued however that within orbifold GUTs it may be possible to satisfy both unification and proton decay bounds with a lower scale \cite{Hall:2001pg}. 

This framework generically leads to a significant component of dark matter in the form of axions. This is expected to be true for both $M$ theory compactifications considered here and the classes of Type IIB compactifications considered in \cite{Bobkov:2010rf}. For the $M$ theory case with $H_I > M_{moduli}$ in particular, when combined with the results of \cite{Acharya:2006ia}, which show that, if stable, there is also a significant component of neutral Wino dark matter, we are led to predict two significant sources of dark matter.

\acknowledgments
B.A. would like to thank R. Akhoury, G. Kane and A. Pierce for useful discussions and thank the MCTP for their hospitaility. P.K. and K.B. would like to thank A. Arvanitaki, S. Dubovsky, B. Dundee and S. Raby for useful discussions.  K.B. is supported by DOE grant DOE/ER/01545-885 and in part by the National Science Foundation under Grant No. PHY05-51164. The work of PK is supported by DOE under contract no. DE-AC02- 05CH11231 and NSF grant PHY-04-57315.

\appendix

\section{Mass Eigenvalues} \label{masses}

In this section, we estimate the mass eigenvalues of the light axions. We will consider the superpotential in  
(\ref{W}) with $N$ geometric axions and one matter axion ($\phi_1$ in the first term in (\ref{W})). 
To that effect, let us define the following $N+1$ linear-independent axionic variables 
\be 
T_K\equiv \chi_1-\chi_{K+1}\,,\,\,\,\,\,K=\overline{1,N+1}\,. 
\ee 
Recall that $T_1$ has already been fixed by (\ref{co}) at $T_1^0=\pi+2\pi n_1$, where $n_1\in{\bf Z}$. 
By examining (\ref{veff}) it is easy to see that in order to minimize the 
potential, the remaining $N$ axions $T_i\,,\,\,\,i>1$ will be fixed  
at the values $T_i^0=2\pi n_i$, where $n_i\in{\bf Z}$. Then, the corresponding  
$N \times N$ Hessian matrix for the light axions 
evaluated at the minimum is approximately diagonal and positive-definite: 
\ba\label{effm} 
\tilde V_{ij}&=&\frac{\partial^2V_{eff}}{\partial T_{i+1}\partial T_{j+1}}\Big |_{T_i=T_i^0}\\ &\approx&\delta_{i+2\,j+2}m_{3/2}e^{K/2}D_{i+2}A_{i+2}e^{-b_{i+2}V_{i+2}}\,,\,\,\,i,j=\overline{1,N}\,.\nonumber\ea 
Possible off-diagonal entries could have come from the terms in (\ref{veff})  
proportional to $e^{-b_kV_k-b_mV_m}$, where $k,m>2$. However, as argued in  
section \ref{stable}, such terms are safe to neglect in the ``double condensate'' regime. 
 
Note that the ``heavy axion'' $T_1$ is an eigenvector of the $N+1\times N+1$ Hessian $V_{AB}$,  
which should in principle be included along with the light axions. 
Clearly, the linear-independent combinations $\{T_K\}$ are eigenvectors of the Hessian.  
However, they do not constitute an orthonormal basis. Indeed, the ``vectors'' $\{T_K\}$  
are related to the ``basis vectors'' $\{t_i,\,\theta\}$ by a general  
$GL(N+1,R)$ linear transformation, which is generally not an orthogonal transformation. 
To construct an orthonormal basis we can use the Gram-Schmidt process as follows: 
\ba\label{grsm} 
&&U_1=T_1\,,\\ 
&&U_2=T_2-\frac{U_1}{U_1\cdot U_1}(U_1\cdot T_2)\,,\nonumber\\ 
&&\,\,\,\,\,\,\,.\,.\,.\,\nonumber\\ 
&&U_{N+1}=T_{N+1}-\sum_{j=1}^{N}\frac{U_j}{U_j\cdot U_j}(U_j\cdot T_{N+1})\,,\nonumber\\ 
&&u_1=\frac{U_1}{|U_1|}\,,u_2=\frac{U_2}{|U_2|}\,,\,.\,.\,.\,,\,u_{N+1}=\frac{U_{N+1}}{|U_{N+1}|}\,.\nonumber 
\ea 
Using the orthonormal basis $\{u_K\}$ we can construct an orthogonal matrix $R\in O(N+1)$ whose $j$-th 
column corresponds to the components of $u_j$ in the basis $\{t_i,\,\theta\}$. 
Hence, the eigenvectors $\{u_K\}$ are related to $\{t_i,\,\theta\}$ by the $O(N+1)$ rotation. 
Let us now compare the eigenvalues of the Hessian corresponding to the $T_K$s with the  
eigenvalues obtained in the orthonormal basis $\{u_K\}$. Note that since the transformation 
between the $T_K$s and the $u_K$s is generally not orthogonal, the corresponding eigenvalues are not 
necessarily the same. Using (\ref{grsm}) one can perform an inverse transformation to  
express the $T_K$s in terms of the orthonormal eigenvectors $\{u_K\}$ 
\ba 
&&T_1=c_1u_1\,,\\ 
&&T_2=c_2u_2+c_{21}u_1\,,\nonumber\\ 
&&T_3=c_3u_3+c_{31}u_1+c_{32}u_2\,,\nonumber\\ 
&&\,\,\,\,\,\,\,.\,.\,.\,\nonumber\\ 
&&T_{N+1}=c_{N+1}u_{N+1}+\sum_{j=1}^Nc_{(N+1)j}u_j\,.\nonumber 
\ea 
From the above, we see an important property 
\be 
\frac{\partial T_K}{\partial u_M}=0\,,\forall M>K 
\,\,\,\Rightarrow\,\,\,\frac{\partial V}{\partial u_K}=c_K\frac{\partial V}{\partial T_K}+\sum_{j=k+1}^{N+1}c_{jK}\frac{\partial V}{\partial T_j}\,. 
\ee 
Furthermore, due to the exponential hierarchy of the terms retained in the sum in (\ref{veff}) we generically expect 
\be 
\frac{\partial V}{\partial T_k}>>\frac{\partial V}{\partial T_{k+1}}\,\,\,\Rightarrow\,\,\,\frac{\partial V}{\partial u_k}\approx c_k\frac{\partial V}{\partial T_k}\,. 
\ee 
Thus, we conclude that  up to some multiplicative factors of $c_k^2\sim {\cal O}(1)$, the eigenvalues  
of the Hessian in the orthonormal basis $\{u_i\}$ are essentially the same as the ones  
in the $\{T_i\}$ basis! Since in this basis the Hessian matrix (\ref{effm}) is approximately diagonal, all we need to 
do is determine the axion decay constants $\hat{f}_i$ by finding a unitary transformation $\tilde{U}_{kl}$, 
which diagonalizes the Kahler metric for the axions and then find the eigenvalues of the matrix: 
\be\label{mmat} 
m_{ij}^2=\frac1{{\hat{f}_{i+2}\hat{f}_{j+2}}}{\tilde U}^{\dagger}_{i m}\tilde{V}_{mn}\tilde{U}_{nj}\,. 
\ee 
In the Kahler metric, the off-diagonal entries are suppressed by a factor ${\tilde a}_i\sim{\cal O}(1/N)$ or ${\tilde a}_i/s_i$ relative to the diagonal ones \cite{Acharya:2008hi}. 
Therefore, treating the Kahler metric as diagonal is justified, which will be confirmed by an explicit numerical example in appendix \ref{toy}. The main effect on the eigenvalues of the Hessian then comes from the rescaling by the inverse decay constants. Thus, a reliable order of magnitude estimate for the masses of the $N$ light axions is given by: 
\ba\label{amass} 
\hat{m}_i^2 &\approx& \frac {m_{3/2}m_{pl}^3}{\hat{f}_{i+2}^2}\,e^{\langle K\rangle/2}c_{i+2}\,e^{-b_{i+2}V_{i+2}};\,i=\overline{1,N}\nonumber\\ 
&\approx& \frac {m_{3/2}m_{pl}^3}{M_{GUT}^2}\,e^{\langle K\rangle/2}c_{i+2}\,e^{-b_{i+2}V_{i+2}}; 
\ea since $\hat{f}_{k} \approx M_{GUT}$. Here $c_i$ is a coefficient of ${\cal O}(1)$. This is the expression used in (\ref{mass-spec}).

\section{An Explicit (toy) Example}\label{toy}

We now study an explicit (toy) example with $N=2$ geometric moduli and a single complex matter field to illustrate the main features of the axion mass spectrum and its decay constants. The K\"{a}hler potential, superpotential and gauge kinetic functions 
are taken as:  
\ba 
K&=&-3 \ln {{\cal V}}+4\pi^{1/3}\frac{\bar\phi_1\phi_1}{{\cal V}};\,\,\,\,\,{\cal V}=4\pi^{1/3}\,s_1^{\frac 76}s_2^{\frac 76}\,, 
\nonumber\\ 
W&=&A_1\phi_1^{-2/P_1}e^{i\frac{2\pi}{P_1}f^1}+A_2e^{i\frac{2\pi}{P_2}f^2}+A_3e^{i\frac{2\pi}{P_3}f^3}\nonumber\\ 
&+&A_4e^{i\frac{2\pi}{P_4}f^4},\nonumber\\ 
f^1&=&f^2=z_1+2z_2;\;f^3=f^4=2z_1+z_2.\nonumber 
\ea 
For the following choice of parameters: 
\ba\label{inpar} 
&&A_1=27\,,\,\,\,A_2=2.27665\,,\,\,\,A_3=3\,,\,\,\,A_4=5\,,\nonumber\\ 
&&P_1=27\,,\,\,\,P_2=30\,,\,\,\,P_3=4\,,\,\,\,P_4=3\,,\nonumber 
\ea 
we obtain 
\ba\label{outpar} 
s_1&\approx& 48.82\,,\,s_2\approx 24.41\,,\,\phi_{1}^0\approx 53.81\,, 
\nonumber\\t_1&\approx& 5\,,\,t_2\approx -10\,,\,\theta_1\approx -15\pi\,. 
\ea 
The geometric moduli $s_1,\, s_2$ and the meson $\phi_{1}^0$ form three mass eigenstates with masses 
\be 
m_1\approx 284.9\,m_{3/2}\,,\,\,m_2\approx 2.0\, m_{3/2},m_3\approx 1.1\, m_{3/2}. 
\ee 
If we define the initial axion basis $(\vec t,\,\vec\theta)=(t_1,t_2,\theta_1)$,  
then the axion kinetic terms are diagonalized by the unitary transformation $U$: 
\be\label{U} 
U\approx\left(\begin{array}{ccc} 
1.00&-10^{-4}&0.01\\ 
10^{-4}&1.00&0.02\\ 
-0.01&-0.02&1.00 
\end{array}\right). 
\ee 
The above matrix indicates that there is very little mixing among the components, which agrees with the remark in section \ref{spectra} and in appendix \ref{masses} that the Kahler metric in the $(\vec t,\,\vec\theta)$ basis is essentially diagonal. From the corresponding eigenvalues we now compute the axion decay constants $f_L\equiv \frac{\hat{f}_L}{m_{pl}}=\sqrt{2{\tilde K}_L}$, labeled in the same order as the eigenvectors corresponding to the columns of $U$ above: 
\be\label{deccon} 
\frac{\hat{f}}{m_{pl}}\approx(3.03\times 10^{-2}\,,\,\,\,6.05\times 10^{-2}\,,\,\,\,1.22)\,. 
\ee 
 
These decay constants are then used to rescale the above eigenvectors 
\be 
U_{KL}\rightarrow U_{KL}{f_L},\,\,{\rm  no\; sum\; over}\, L\,, 
\ee 
to obtain canonical kinetic terms for the axions. It is possible to retain good accuracy by simply 
treating the Kahler metric as diagonal, in which case the axion decay constants can be extracted from the 
diagonal components of the Kahler metric as:  
\be\label{dcon} 
f_i\approx\sqrt{2{\tilde K}_{ii}},\;\;f_{\theta_k}\approx\sqrt{2{\tilde K}_{\theta_k\theta_k}} 
\ee 
The decay constants in (\ref{deccon}) are in fairly good agreement with analytical estimate above. 
 
The next step is to determine the unitary transformation $\cal U$ that diagonalizes the mass matrix (\ref{mass-matrix}). 
It is given by: 
\be\label{Ucal} 
{\cal U}\approx\left(\begin{array}{cccc} 
0.706&0.708&-0.019\\ 
0.706&-0.702&0.093\\ 
-0.053&0.079&0.995 
\end{array}\right). 
\ee 
It is convenient to express all the masses relative to the gravitino mass scale $m_{3/2}$. The axion masses obtained from diagonalizing the mass matrix (\ref{mass-matrix}) are: 
\ba\label{anumer} 
&&\hat{m}_{\psi_1}\approx 286\,m_{3/2}\,,\,\,\,\,\hat{m}_{\psi_2}\approx 6.3\times 10^{-35}\,m_{3/2}\,,\\ 
&&\hat{m}_{\psi_3}\approx 4.0\times 10^{-51}\,m_{3/2}.\nonumber 
\ea 
Expressing the above masses in units of Planck mass we obtain: 
\ba\label{mampl} 
&&\hat{m}^2_{\psi_1}\approx 1.1\times 10^{-27}\,m_{p}^2\,,\,\,\hat{m}^2_{\psi_2}\approx 5.2\times 10^{-101}\,m_{p}^2\,,\nonumber\\ 
&&\hat{m}^2_{\psi_3}\approx 2.1\times 10^{-133}\,m_{p}^2\,, 
\ea 
where we used the value of the gravitino mass specific to the above numerical example\footnote{Such a low gravitino mass scale 
is an artifact of the toy model that has only two moduli. In this case the seven dimensional volume $V_X=s_1^{7/6}s_2^{7/6}\approx 3880$ is rather large, which makes the gravitino mass smaller than in the more 
realistic examples where $V_X\sim{\cal O}(500-1000)$.} $m_{3/2}\approx 277$ GeV. 
We can now compare the masses of the axions obtained numerically with the approximate analytical formula (\ref{amass}).  
 
Using the general formula (\ref{amass}), and using the expression for $f_{i+2}$ in (\ref{dcon}) with $c_{i+2}=1$, we obtain an order of magnitude estimate for the axion masses: 
\ba 
&&\hat{m}_{\psi_1}\approx 68\,m_{3/2}\,,\,\,\hat{m}_{\psi_2}\approx 3\times 10^{-36}\,m_{3/2}\,,\nonumber\\ 
&&\hat{m}_{\psi_3}\approx 6\times 10^{-50}m_{3/2}\,, 
\ea 
which upon comparing with the exact numerical result (\ref{anumer}),  
confirms that (\ref{amass}) is a valid approximation for masses of both light and heavy mass eigenstates. 
 
Let us express the original axion fields $(\vec t,\,\vec\theta)$ in terms of the canonically normalized mass eigenstates (before taking QCD effects into account): 
\ba 
&&t_1=23.3\psi_1+23.4\psi_2-0.6\psi_3\,,\\ 
&&t_2=11.7\psi_1-11.6\psi_2+1.6\psi_3\,,\nonumber\\ 
&&\theta_1=-0.6\psi_1+6.3\times 10^{-2}\psi_2+0.8\psi_3\,.\nonumber 
\ea 
The effective decay constants ${\tilde f}_K$ for the mass eigenstates $\psi_K$ can be computed using the general formula (\ref{feff}), where index $K$ now runs over $1\leq K<N+1$. Keeping the coefficients ${\tilde N}_i^{\rm vis}$ arbitrary, the effective decay constants for our toy example are given by: 
 
\ba\label{feff1} 
&&{\tilde f}_1=\left(146.5\times {\tilde N}^{\rm vis}_1+73.2\times {\tilde N}^{\rm vis}_2\right)^{-1}m_{p}\,,\\ 
&&{\tilde f}_2=\left(146.9\times {\tilde N}^{\rm vis}_1-72.8\times {\tilde N}^{\rm vis}_2\right)^{-1}m_{p}\,,\nonumber\\ 
&&{\tilde f}_3=\left(3.8\times {\tilde N}^{\rm vis}_1+9.8\times {\tilde N}^{\rm vis}_2\right)^{-1}m_{p}\,,\nonumber 
\ea 
where we took into account the factor of $2\pi$ multiplying the fields $t_i$. From the arguments in section \ref{qcdaxion}, it is clear that the lightest eigenstates $\psi_2$ and $\psi_3$ with $\hat{m}_{\psi_2}^2\approx 5.2\times 10^{-101}\,m_{pl}^2$ and $\hat{m}_{\psi_3}^2\approx 2.1\times 10^{-133}\,m_{pl}^2$, which are much smaller than $m_{exp}^2=10^{-82}\,m_{pl}^2$ (see eqn.(\ref{mexp})), and with effective decay constants $\tilde{f}_i\sim{\cal O}(M_{\rm GUT})$, are good candidates for the QCD axion. As explained in section \ref{qcdaxion}, the QCD axion is in general a linear combination of $t_i$'s, for simplicity we choose $\tilde{N}_1^{\rm vis}=\tilde{N}_2^{\rm vis}=1$. This linear combination representing the QCD axion can be written in terms of the mass eigenstates, before the QCD instanton effects are taken into account: 
\ba 
\theta_{qcd} &=& 2\pi\left(\tilde{N}_1^{\rm vis}t_1+\tilde{N}_2^{\rm vis}t_2\right)\\ 
&=&2\pi(t_1+t_2) 
\approx 219.8\psi_1+74.1\psi_2+5.9\psi_3\,,\nonumber 
\ea 
 
Finally, we take into account QCD instanton effects and determine precisely which axion candidate satisfies all criteria for being the QCD axion. The linear combination of \emph{final} mass eigenstates, which directly couples to the visible sector is given by: 
\ba\label{lqcd} 
\theta_{qcd}&=&2\pi(\tilde{N}_1^{\rm vis}t_1+\tilde{N}_2^{\rm vis}t_2)=2\pi(t_1+t_2) \\ 
&\approx& 219.8\,{\tilde\psi}_1+5.5\times 10^{-28}{\tilde\psi}_2-74.3\,{\tilde\psi}_3.\nonumber 
\ea 
and the \emph{final} mass spectrum is given by: 
\ba 
{\hat m}^2_{\tilde{\psi}_1}&\approx& 1.1\times 10^{-27} m_{p}^2\,,\,\,\,{\hat m}^2_{\tilde{\psi}_2}\approx 3.3\times 10^{-103} m_{p}^2\,,\nonumber\\ 
{\hat m}^2_{\tilde{\psi}_3}&\approx& 5.5\times 10^{-73} m_{p}^2\,. 
\ea 
Note that according to the arguments in section \ref{qcdaxion}, the mass of the heavy axion eigenstate ${\tilde{\psi}_1}$ is the same as that for ${\psi}_1$ from (\ref{mampl})  
while the mass of lightest eigenstate ${\psi}_3$ in (\ref{mampl}) is completely modified, receiving a mass predominantly from QCD instanton effects. The mass of  
${\psi}_2$ also receives a noticeable modification from QCD instantons. The eigenstates $\tilde{\psi}_2$ and $\tilde{\psi}_3$ are therefore modified from $\psi_2$  
and $\psi_3$ respectively. $\tilde{\psi_3}$ can be identified with the QCD axion, which agrees with the expectation from the arguments in section \ref{qcdaxion}. 
 
The eigenstates with significant couplings to the visible sector are the heavy state ${\tilde \psi}_1$, and the light eigenstate ${\tilde \psi}_3$ that picks up its mass from the QCD instanton effects and is of $m_{a}^{qcd}={\cal O}(\Lambda_{qcd}^2/f_a)$. In more general cases, there are generically other eigenstates with masses heavier than $m_{a}^{qcd}$, so they will also couple appreciably to the visible sector, in particular to $\vec{E}\cdot\vec{B}$. On the other hand, eigenstates much lighter than $m_{a}^{qcd}$ ($\tilde{\psi}_2$ in (\ref{lqcd}) above) will \emph{not} couple appreciably to $\vec{E}\cdot\vec{B}$, as their couplings to the visible sector are expected to be suppressed by the mass ratio $({\hat m}_{{\tilde \psi}_k}/\hat{m}_a^{qcd})^2$. From (\ref{lqcd}) we see that this is indeed the case for $\tilde{\psi}_2$. 

\section{Axion Decay Constants}\label{stats}

\begin{figure}[h!]\label{aidist}
 \begin{center}
\leavevmode \epsfxsize 7.5 cm \epsfbox{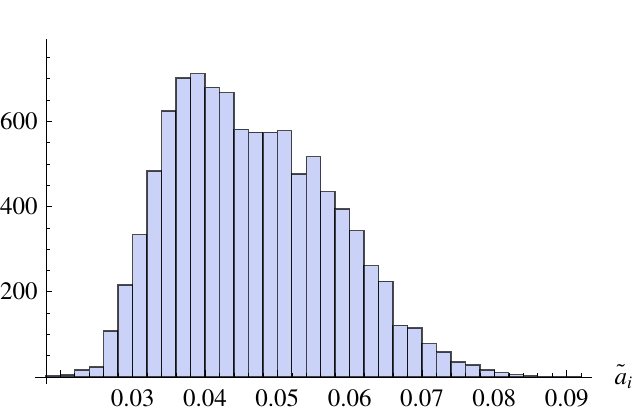}
\end{center}
 \caption{Distribution of ${\tilde a}_i$ obtained for $200$ randomly generated Kahler potentials consistent with $G_2$ holonomy 
with $N=50$ moduli. The for each case, the integer parameters $N_i$ were randomly generated sets containing $1$s and $2$s. The mean value is 
${\overline {\tilde a}}_i=\frac 7{150}\approx 0.047$ and the standard deviation ${\rm S.D.}({{\tilde a}_i})\approx 0.011$.}
\end{figure}

In this section, we compute the generic spectrum of axion decay constants $\hat{f}_a$, as well as effective decay constants $\tilde{f}_a$ that are relevant in determining the axion couplings to the visible sector, in a framework in which all moduli and axions are stabilized from a set of microscopic ``data". Instead of relying on any specific choice of the $G_2$ manifold, we will use some of the generic properties of the Kahler metric for the axions and the moduli to give  
an order of magnitude estimate. We perform a simple statistical analysis of the axion decay constants by considering a general class of Kahler potentials consistent with $G_2$ holonomy where the seven-dimensional volume is given by 
\be 
{\cal V}=4\pi^{1/3}\,\sum_{k=1}^M c_k\prod_{i}s_i^{{a}_i^k}\,,\,\,\,\,\,\,\,\sum_{i=1}^N{a}_i^k=7/3\,,\,\,\,\,\forall k\,, 
\ee 
where $c_k\sim{\cal O}(1)$ are integer coefficients and the exponents $\frac 13\leq {a}_i^k\leq \frac 73$ are multiples of $\frac 13$ so that each product term in the sum contains a maximum of seven distinct factors. The latter condition was motivated by the form of the Kahler potential for Joyce orbifolds and imposed for simplicity but may be relaxed by considering smaller values of ${a}_i^k$. We assume that a manifold has $N=50$ moduli and that the number of distinct terms inside the sum is $M\sim{\cal O}(1000)$. 
To determine the moduli vevs at the minimum we first need to solve the system of equations  
to determine the parameters ${\tilde a}_i$, as explained in \cite{Acharya:2008hi}. Here it is also assumed for simplicity that the integers $N_i$ inside the gauge  
kinetic function of the dominant gaugino condensates are random sets containing $1$'s and $2$'s. With these inputs, the distribution of ${\tilde a}_i$ is presented in Figure \ref{aidist}, which is rather broad. Such broadening can be attributed to the variation of the integers $N_i$ since increasing a particular $N_i$ leads to a slight decrease in the value of the corresponding ${\tilde a}_i$, as observed from numerical simulations. The distribution of ${\tilde a}_i$ in Figure \ref{aidist} was generated by solving a system of equations \cite{Acharya:2008hi} for $200$ distinct randomly generated Kahler potentials of the form described above with $N=50$ moduli. For each solution we have verified that all $50$ parameters ${\tilde a}_i$ always add up to $7/3$, as expected. The mean and thermal deviation are given by: 
\be 
{\overline {\tilde a}}_i=\frac 7{150}\approx 0.047,\;{\rm S.D.}({{\tilde a}_i})\approx 0.011\,. 
\ee 
 
\begin{figure}[h!]\label{effdec}
 \begin{center}
\leavevmode \epsfxsize 7.5 cm \epsfbox{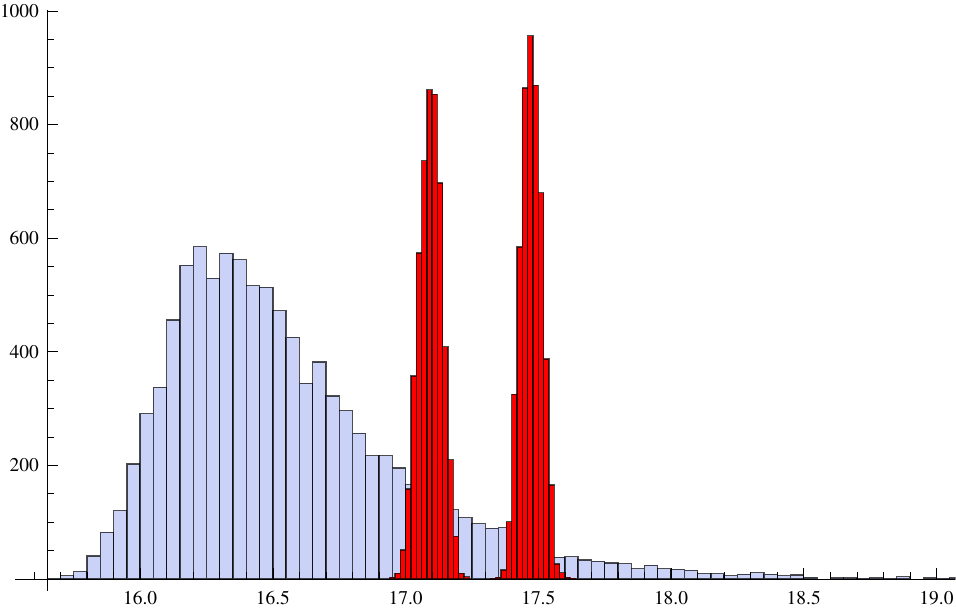}
\end{center}
 \caption{The broad distribution in light blue color corresponds to absolute values of the effective decay constants on the logarithmic scale ${\rm Log}_{10}(|{\tilde f}_L|/{\rm GeV})$ 
whereas the double-peaked distribution in red color corresponds to the original decay constants ${\rm Log}_{10}(\hat{f}_L/{\rm GeV})$ on the same scale.}
\end{figure}

We now use the same $200$ sets of parameters ${\tilde a}_i$ along with the corresponding integers $N_i$
to compute the moduli vevs at the minimum by following the analysis in \cite{Acharya:2008hi}, where we choose $Q=60$ and $Q-P=3$, and then use those to evaluate the Kahler metric at the minimum. The Kahler metric ${\tilde K}_{AB}$ is diagonalized for each set with all eigenvalues strictly positive. Note that we have not included the meson Kahler metric into our analysis. The corresponding decay constant is an order of magnutude larger that the axion decay costants corresponding to $t_i$s, hence its inverse gives the smallest contribution when computing the effective decay constants as will be seen later. We then compute the axion decay constants by using the definition $\hat{f}_L=\sqrt{2\tilde{K}_L}\,m_p$. Two sharp peaks colored in red in Figure \ref{effdec} represent the distribution of the
axion decay constants on a logarithmic scale. The double-peak shape can be traced back to the fact that there are two distinct sets of moduli vevs. The peaks correspond to the following values of the decay constants:
\ba
\left(f_L\right)_{{\rm Peak}_1}\approx 1.3\times 10^{17}\,{\rm GeV},\,
\left(f_L\right)_{{\rm Peak}_2}\approx 2.9\times 10^{17}\,{\rm GeV}\nonumber
\ea 
The above result can also be obtained by a simple approximation. Neglecting the difference due to the integers $N_i$ and using the diagonal components of the Kahler metric $\tilde{K}_{ab}\equiv \frac{\partial^2\,K}{\partial w_a\partial w_b};w\equiv\{t_i,\theta_j\}$, results in the following parametric dependence of the axion decay constants on $N$ and $Q$:
\be \label{fapprox}
\hat{f}_L\approx 1.6\frac{\sqrt N}{Q}\times 10^{18}\,{\rm GeV}\,,
\ee
which for $Q=60$ and $N=50$ results in $\hat{f}_L\approx 1.8\times 10^{17}\,{\rm GeV}$, very close to the values obtained by the statistical method above. It is clear that by keeping the number of moduli $N$ fixed while increasing 
the dual Coxeter number $Q$ of the gauge group, one can lower the axion decay constants. Thus, it seems reasonable to obtain axion decay constants of magnitude $10^{16}$-$10^{17}$ GeV, consistent with standard gauge unification.

Finally, we use eqn. (\ref{feff}) to compute the effective decay constants, where for each set of $\hat{f}_L$s,
the unitary matrix ${\cal U}_{KL}$ used in (\ref{feff}) was obtained by diagonalizing a randomly generated symmetric matrix, while $U_{KL}$ was the actual unitary transformation that diagonalized the Kahler metric. Here it is assumed that the integers $N^{\rm vis}_i$ of the visible sector gauge kinetic function are randomly generated sets containing $0,\,1,\,2$. 
The distribution of absolute values of the effective decay constants on a logarithmic scale is presented in Figure \ref{effdec} in light blue color. Since the distribution is clearly non-Gaussian, the peak value is somewhat smaller than the mean value:
\ba
\left(|{\tilde f}_L|\right)_{\rm Peak}\approx 1.6\times 10^{16}\,{\rm GeV},\,\left(|{\tilde f}_L|\right)_{\rm mean}\approx 3.6\times 10^{16}\,{\rm GeV}\nonumber
\ea
Thus, from the above analysis the value of the effective decay constants are expected to be few $\times 10^{16}$ GeV, the same as the scale of standard gauge unification.

\end{document}